\documentclass[pdflatex,sn-nature]{sn-jnl}

\usepackage{graphicx}
\usepackage{amsmath,amssymb}
\usepackage{booktabs}
\usepackage{upgreek}
\usepackage{subcaption}
\usepackage{textcomp}

\usepackage{hyperref}
\usepackage{setspace}
\usepackage{geometry}
\title[Braided endovascular implants for intracranial aneurysms]{Braided endovascular implants for intracranial aneurysms: mechanics, hemodynamics, and clinical translation}

\author*[1]{\fnm{Ratnadeep} \sur{Pramanik}}
\email{ratnadeep.pramanik@unibw.de}
\author[2,3]{\fnm{Duygu} \sur{Dengiz}}
\email{dude@tf.uni-kiel.de}
\author[4]{\fnm{Mariya S.} \sur{Pravdivtseva}}
\email{mariya.pravdivtseva@rad.uni-kiel.de}
\author[1]{\fnm{Martin} \sur{Frank}}
\email{martin.frank@unibw.de}
\author[1]{\fnm{Ivo} \sur{Steinbrecher}}
\email{ivo.steinbrecher@unibw.de}
\author[5]{\fnm{Prasanth} \sur{Velvaluri}}
\email{prasanth.velvaluri@imtek.uni-freiburg.de}
\author[1,6]{\fnm{Matthias} \sur{Mayr}}
\email{matthias.mayr@unibw.de}
\author[7]{\fnm{Philipp} \sur{Berg}}
\email{philipp.berg@ovgu.de}
\author[8]{\fnm{Sylvia} \sur{Saalfeld}}
\email{saalfeld@medinfo.uni-kiel.de}
\author[3]{\fnm{Naomi} \sur{Larsen}}
\email{Naomi.Larsen@uksh.de}
\author[3]{\fnm{Olav} \sur{Jansen}}
\email{olav.jansen@uksh.de}
\author[1]{\fnm{Alexander} \sur{Popp}}
\email{alexander.popp@unibw.de}

\affil[1]{Institute for Mathematics and Computer-Based Simulation, University of the Bundeswehr Munich, 85577 Neubiberg, Germany}
\affil[2]{Materials Science and Mechanical Engineering, Harvard John A. Paulson School of Engineering and Applied Sciences, Harvard University, 02134 Allston, MA, USA}
\affil[3]{Department of Radiology and Neuroradiology, University Hospital Schleswig-Holstein, Kiel University, 24105 Kiel, Germany}
\affil[4]{VIVID Research Group, Department of Radiology and Neuroradiology, University Hospital Schleswig-Holstein, Kiel University, 24105 Kiel, Germany}
\affil[5]{Institute of Microsystems Engineering (IMTEK), IMBIT//NeuroProbes, BrainLinks-BrainTools, University of Freiburg, 79110 Freiburg, Germany}
\affil[6]{Data Science \& Computing Lab, University of the Bundeswehr Munich, 85577 Neubiberg, Germany}
\affil[7]{Department of Medical Engineering, Otto von Guericke University Magdeburg, 39106 Magdeburg, Germany}
\affil[8]{Institute for Medical Informatics and Statistics, Kiel University and University Hospital Schleswig-Holstein, 24105 Kiel, Germany}

\abstract{Endovascular implants prevent intracranial aneurysm rupture by altering the mechanical and hemodynamic environment at the aneurysm neck. Yet many \textit{in silico} workflows prescribe or reconstruct the post-deployment geometry before computing flow, leaving unresolved the mechanics that create the clinically relevant interface. Here we review braided intraluminal flow diverters, intrasaccular devices, and emerging flow-disruption concepts across deployment mechanics, inter-wire and wire-wall contact, superelasticity, wall apposition, pore geometry, computational fluid dynamics, and fluid-structure interaction. We connect these modeling choices to neck coverage, malapposition, migration, deformation, and durability, and distinguish established evidence from mechanistic inference and prospective hypotheses. We argue that model fidelity should match the clinical question: prescribed or fast placement may support screening, whereas questions of coverage, apposition, compaction, and migration benefit from mechanically plausible deployment states. An interface-resolved mechanics-to-flow framework, supported by measurable validation targets and standardized reporting, could improve device design and enable more reliable patient-specific treatment planning.}


\begin{document}
\maketitle

\section{Introduction}
\label{sec:introduction}

Intracranial aneurysms are abnormal focal dilations of cerebral arteries and affect about 3\% of adults~\cite {etminan2016unruptured,backes2016patient}. Aneurysm rupture causes subarachnoid haemorrhage, a catastrophic form of stroke with high morbidity and mortality~\cite{texakalidis2019aneurysm,brisman2006cerebral}. Clinical management of an unruptured aneurysm therefore requires a careful balance between its long-term risk of rupture and the hazards of intervention. This balance depends on patient characteristics, aneurysm morphology and location, treatment strategy, and neurointerventional expertise.

Endovascular interventions seek to isolate the aneurysm sac from the parent circulation and promote thrombosis, endothelialization, and stable occlusion. Current strategies include detachable coils, balloon- or stent-assisted coiling, intraluminal flow diverters (FDs), and intrasaccular flow-disruption devices~\cite{goubergrits2014hemodynamic,pierot2013endovascular,lauzier2023review}. FDs bridge the aneurysm neck within the parent artery and provide a scaffold for vessel-wall reconstruction. Intrasaccular devices, such as the Woven EndoBridge (WEB) and the Contour Neurovascular System (CNS), occupy the aneurysm sac and disrupt flow at the ostium~\cite{muskens2017woven,heiferman2024new,akhunbay2020endovascular,zhuo2025intrasaccular}. Although their deployment strategies differ, both device families rely on the structural configuration established at the implant-vasculature interface.

Inadequately sized or poorly positioned implants can leave parts of the aneurysm neck exposed, create malapposition gaps, migrate, compact, or deform. For FDs, microcatheter-induced twisting, secondary collapse, incomplete wall apposition, branch-vessel jailing, and distal migration can limit treatment efficacy. For intrasaccular devices, insufficient neck coverage, inaccurate sizing, post-deployment deformation, and delayed compaction are associated with recurrence~\cite{zhou2017complications,velvaluri2021torsional,hohenstatt2020branch,muhlbenninghaus2019transient}. These problems motivate predictive \textit{in silico} modeling. A clinically useful computational workflow should compare biomechanically plausible device configurations and help identify the implant type and size most likely to provide stable neck coverage and meaningful flow attenuation.

These failure modes are often discussed separately, yet they originate from the same local problem. The device must first achieve a stable configuration within curved, deformable anatomy; that configuration then sets the pores through which blood enters the sac. A small gap may have little effect on global porosity but still form a concentrated inflow channel. Likewise, two devices with the same nominal diameter can exhibit different local coverage due to foreshortening or compression. This is why nominal device specifications alone are insufficient for patient-specific interpretation.

\begin{figure}[htpb!]
\centering
\includegraphics[width=\textwidth]{relevance.jpg}
\caption{Descriptive publication trends from 1990 to 2025 for PubMed-indexed studies on intracranial aneurysms and endovascular implants. Annual counts were obtained from PubMed (\url{https://pubmed.ncbi.nlm.nih.gov/}) using \textit{intracranial aneurysm} for the broader literature and the implant-related terms \textit{endovascular device}, \textit{neurovascular device}, \textit{endovascular implants}, and \textit{neurovascular implants}. The two literatures are shown on separate vertical axes; the comparison provides context for the present review and is not a study-selection flow diagram.}
\label{fig:publication_trends}
\end{figure}

Patient-specific modeling begins with clinical imaging. Magnetic resonance angiography, computed tomography angiography, digital subtraction angiography (DSA), and three-dimensional rotational angiography (3D\,RA) provide complementary anatomical information for treatment planning~\cite{howard2019comprehensive}. DSA and 3D\,RA remain central to neurointerventional planning because they combine high spatial resolution with procedural information. For \textit{in silico} workflows, 3D\,RA is particularly useful for reconstructing the aneurysm neck and nearby branch origins.

Vascular virtual anatomies establish the domain for virtual deployment and blood flow simulation. Typical pipelines segment vasculature from the medical images, extract surface meshes, smooth or repair the surface, and truncate branches that are unlikely to affect local aneurysm hemodynamics~\cite{adams1994seeded,lorensen1987marching,field1988laplacian}. The segmentation approaches use intensity thresholding, level-set methods, deformable models, manual curation, centerline-guided editing, or machine learning. Whatever the method, the reconstructed neck, parent vessel, and adjacent branches must retain the anatomical features that determine device position and flow.

Geometric uncertainty is also spatially uneven. A small surface change in the aneurysm dome may have limited influence on deployment. In contrast, a similar change at the neck rim, a landing zone, or a branch origin may alter position and coverage substantially. Validation should therefore examine local anatomical fidelity at the implant interface rather than relying only on a global surface-distance measure. The same principle applies to smoothing: removing imaging noise is useful, but excessive smoothing can erase precisely the curvature or neck feature that provides mechanical support.

This focused review synthesizes the structural, hemodynamic, and translational literature on braided endovascular implants for intracranial aneurysms. We identified the literature through targeted PubMed searches and reference-list screening using combinations of terms related to intracranial aneurysms, flow diversion, intrasaccular devices, braided implants, virtual deployment, contact mechanics, computational fluid dynamics (CFD), fluid-structure interaction (FSI), and validation. The aim is conceptual integration rather than systematic evidence pooling; no formal risk-of-bias assessment or meta-analysis was undertaken. We therefore distinguish established findings supported by experimental, imaging, or clinical evidence from mechanistic inferences based mainly on computation or bench testing, and from prospective hypotheses that still require validation.

\begin{figure}[htpb!]
\centering
\includegraphics[width=\textwidth]{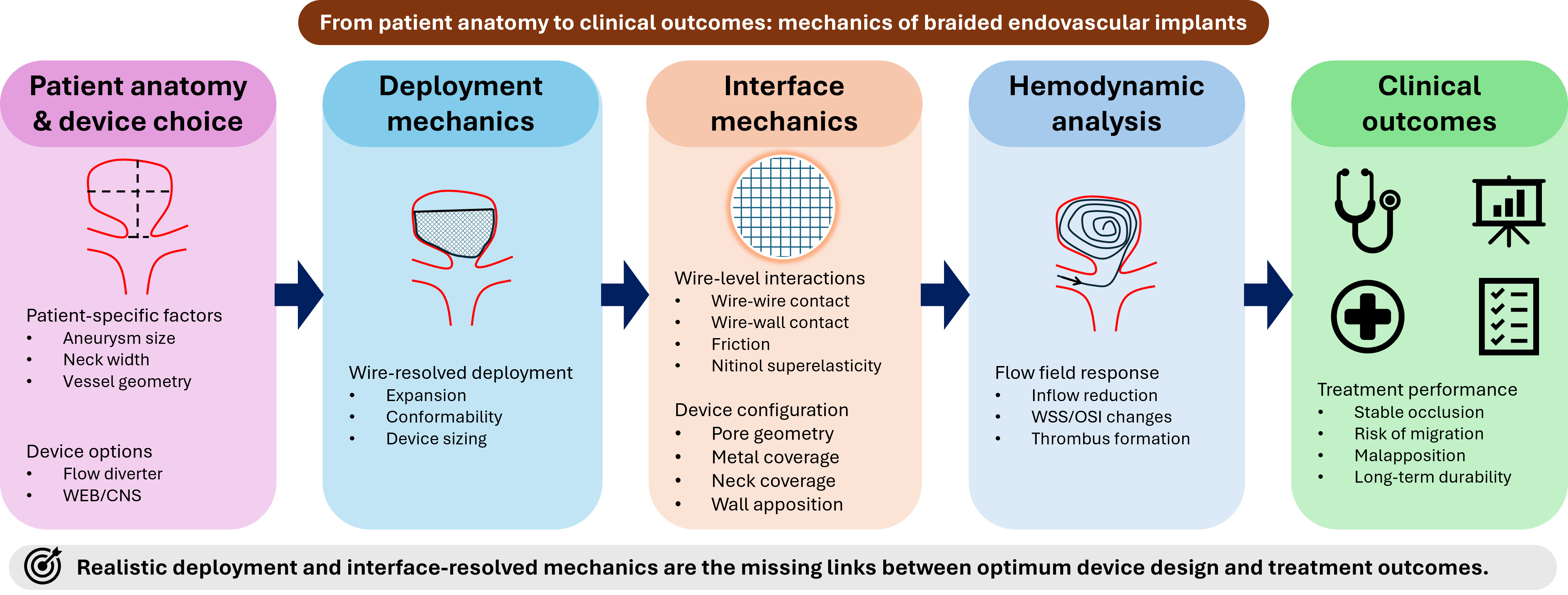}
\caption{Mechanics-to-flow framework for braided endovascular implants. Patient anatomy and device design define the deployment problem; contact, friction, and boundary constraints establish the realized braid geometry, which sets the neck-plane pore structure used for hemodynamic analysis and clinical interpretation.}
\label{fig:graphical-abstract}
\end{figure}

Publication activity provides additional context for this scope. As shown in Fig.~\ref{fig:publication_trends}, both the broader intracranial aneurysm literature and the more focused endovascular-implant literature expanded markedly through 2025, although the implant-specific evidence base remains much smaller. These descriptive counts were used only to contextualize the field and did not define study eligibility for this review. The review follows the mechanics-to-flow sequence shown in Fig.~\ref{fig:graphical-abstract}. Image-based anatomy and device specifications define the deployment problem; deployment establishes pore geometry, neck coverage, and apposition; and that realized configuration becomes the domain for hemodynamic analysis. The post-deployment device is therefore not a neutral preprocessing choice. Its physical fidelity sets the limit on what can reasonably be inferred from the subsequent flow calculation.

Our objective is to evaluate where current modeling strategies capture, simplify, or omit the interfaces that matter for clinical interpretation. We focus on FDs, WEB, CNS, and related emerging flow disruptors, and connect multi-body mechanics to downstream fluid transport, durability, and device selection. The result is an evidence-informed roadmap for matching computational fidelity to the clinical question and for reporting patient-specific simulations reproducibly. Aneurysm biology, thrombus maturation, drug-eluting coatings, inflammatory cascades, peridevice edema, and chronic neointimal growth are considered only where they define the limits of a mechanical or hemodynamic interpretation.

\section{Endovascular implant architectures}
\label{sec:devices}

\subsection{Established devices}
\label{subsec:established_device_families}

\begin{figure}[htpb!]
\centering
\includegraphics[width=\textwidth]{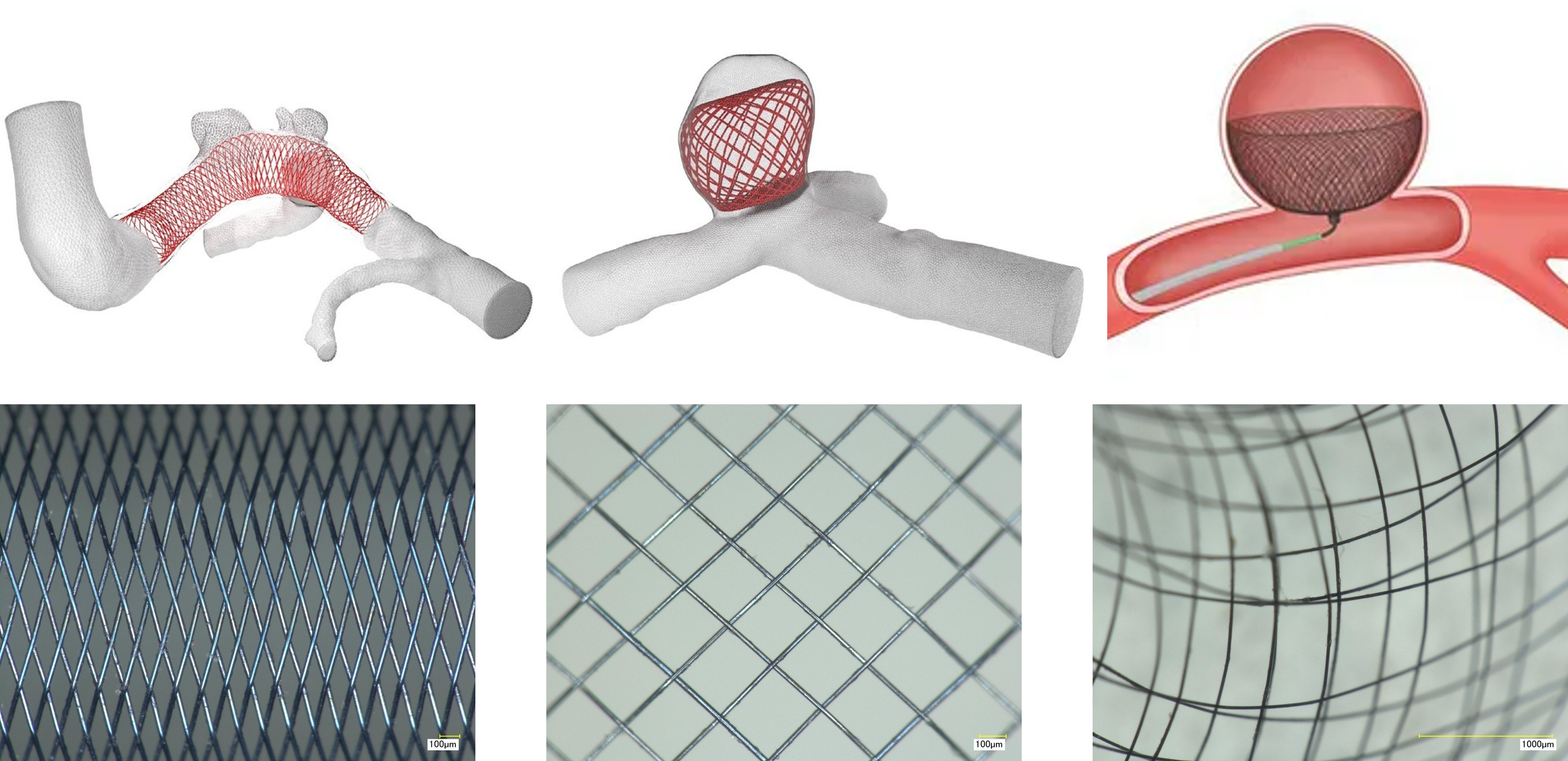}
\caption{Representative configurations (top) and zoomed-in optical microscopic images (bottom) of braided endovascular implants. An intraluminal FD (left) screens the aneurysm neck from the parent vessel, whereas the WEB device (middle) and CNS (right) disrupt flow from within the aneurysm sac. (top left, top middle) Reproduced from previous work by our co-author~\cite{frank2024numerical}, and (top right) reproduced with permission from~\cite{zhuo2025intrasaccular}.}
\label{fig:beds_braids}
\end{figure}

The main device families occupy different anatomical interfaces (Fig.~\ref{fig:beds_braids}). FDs are deployed across the aneurysm neck within the parent vessel. Their braided mesh reduces inflow into the sac while providing a scaffold for endothelial growth across the neck. The immediate effect is hemodynamic: the mesh weakens the inflow jet and promotes regions of slow flow. Stable occlusion then depends on thrombus organization and subsequent tissue growth.

By contrast, WEB and CNS devices are deployed at the aneurysm ostium and proximal sac. Their dense braided meshes interrupt incoming momentum at the neck and provide a local scaffold for thrombosis and tissue coverage. For FDs, modeled performance depends on landing-zone morphology, foreshortening, wall apposition, branch-vessel coverage, and the local metal coverage ratio (MCR)~\cite{kallmes2007new,kim2024quantitative}. For intrasaccular implants, the corresponding variables are device size, proximal and distal pole position, ostium coverage, uncovered neck area, anatomical support, and time-dependent deformation~\cite{mantilla2023woven,monteiro2022treatment}.

Sizing errors have different consequences across these device classes. FD oversizing changes the braid angle, porosity, and deployed length, and may increase branch-vessel coverage; undersizing can compromise apposition and promote migration. For an intrasaccular device, undersizing can leave part of the ostium exposed, whereas oversizing improves anchoring at the cost of greater compression, deformation, and possibly higher local wall stress. In small aneurysms ($<10$\,mm), the challenge is stable ostial coverage without compromising adjacent branches. In giant aneurysms ($\ge 20$--$25$\,mm), mechanics are more closely entangled with thrombus organization, inflammation, and peridevice edema~\cite{texakalidis2019aneurysm,soldozy2019biophysical,ngoepe2018thrombosis}. These biological processes are outside the main scope of this review, but they limit how far a purely mechanical model can be interpreted clinically.

The relevant interface metric also changes with device class. For an FD, circumferential wall contact and the spatial distribution of MCR along the landing zone matter in addition to coverage at the neck. For an intrasaccular device, the uncovered ostium area and support provided by the aneurysm rim are more direct measures. Reporting a single global porosity or compression ratio can obscure these differences. Device-specific metrics are therefore needed if simulations are to compare treatment options rather than only describe one deployment.

\subsection{Emerging concepts}
\label{subsec:emerging_implants}

Next-generation implants optimize the device-vasculature interface through innovative anchoring, geometric adjustments, or targeted ostium shielding. For instance, a newly developed flow disruptor merges intraluminal diversion with intrasaccular flow disruption~\cite{won2025efficacy}. Initial \textit{in silico} and bench testing emphasizes that flow attenuation remains highly sensitive to the final deployed configuration and local boundary constraints. Alternative intrasaccular and neck-bridging options include the Artisse and SEAL-type platforms~\cite{hecker2025artisse,hecker2026artisse,zoppo2024novel}. Shifting away from traditional weaves, thin-film helical intravascular structures represent another evolving paradigm. A patient-specific transient CFD analysis compared a helical thin-film nitinol device against a standard braided FD~\cite{Voss2025}. The helical design delivered equivalent or superior performance, yielding mean hemodynamic reductions of 52--58\% compared to 47\% for the FD. Peak flow reductions reached 70\% with shorter variants, though the device demonstrated a higher sensitivity to deployment positioning and spatial orientation.

\begin{figure}[htpb!]
\centering
\includegraphics[width=\textwidth]{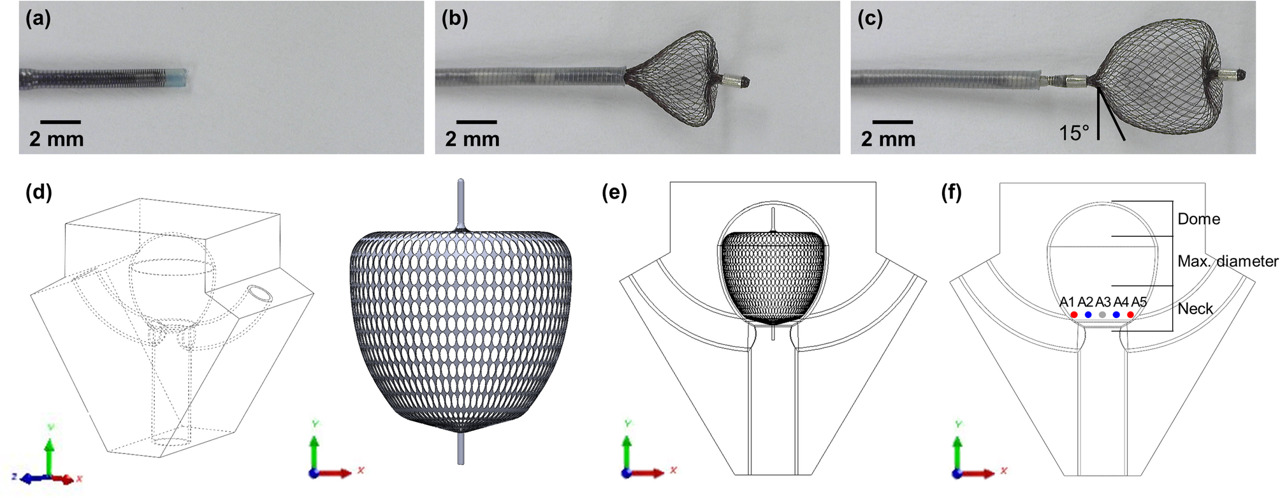}
\caption{Bench-top microcatheter deployment and three-dimensional computational modeling of a novel flow disruptor (NFD). Panels (\textbf{a}--\textbf{c}) show the NFD loaded into a 4.2~Fr intermediate catheter using manual crimping, during partial release, and after full deployment into its characteristic conical configuration. Panels (\textbf{d}--\textbf{f}) show the bifurcation-aneurysm model, the reconstructed and virtually deployed NFD, and the subregions A1--A5 used for localized velocity measurements within the aneurysm sac. Reproduced with permission from~\cite{won2025efficacy}.}
\label{fig:nfd_deployment}
\end{figure}

\begin{figure}[htpb!]
\centering
\includegraphics[width=\textwidth]{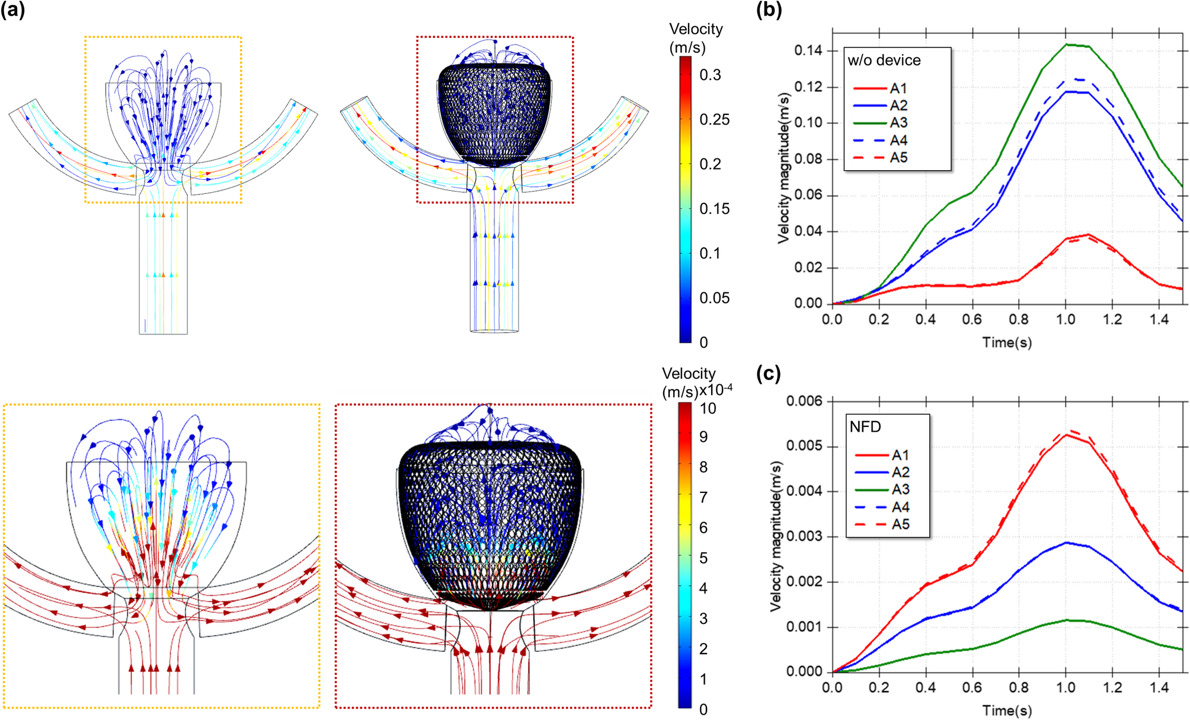}
\caption{Hemodynamic assessment of a bifurcation-aneurysm model before and after NFD placement. Velocity streamlines and the corresponding regional velocity histories illustrate how the realized device configuration attenuates intra-aneurysmal flow. Reproduced with permission from~\cite{won2025efficacy}.}
\label{fig:nfd_hemodynamics}
\end{figure}

The mechanics-to-flow link is illustrated directly in Figs.~\ref{fig:nfd_deployment} and~\ref{fig:nfd_hemodynamics}. The first juxtaposes microcatheter loading and release with the reconstructed deployment state and local measurement regions, whereas the second compares pre- and post-deployment velocity fields and regional velocity histories. Together, they show why the mechanically realized device shape and boundary contact should be treated as inputs to the hemodynamic model rather than as incidental visualization.

Concurrently, hemispherical NiTi thin-film implants utilize a comparable manufacturing approach but are tailored for intrasaccular placement at the aneurysm neck~\cite{dengiz2025thin,dengiz2026mechanical}. These structures incorporate a structural backbone (42~\textmu m) alongside ultra-thin mobile flaps (7~\textmu m). This distinct architectural feature permits post-operative access to the aneurysm sac, preserving options for secondary endovascular interventions if complications arise. Comprehensive mechanical benchmarking across several variations demonstrated up to a 52\% increase in radial force relative to existing commercial options, though further experimental fluid-dynamics characterization is still required. Ultimately, local anatomical constraints dictate the operational braid angle, pore distribution, wall contact, and neck coverage. The clinically relevant device state is therefore a coupled function of implant architecture, patient-specific anatomy, material response, and deployment mechanics.

Table~\ref{tab:device_model_map} links each device class to its principal interface questions, modeling options, and clinically interpretable outputs. The appropriate level of simulation ranges from rapid geometric placement for screening to contact-resolved deployment and coupled multiphysics when the question depends on local mechanics.

\begin{table*}[htpb!]
\centering
\caption{Device-specific interface questions, modeling strategies, and clinically relevant outputs.}
\label{tab:device_model_map}
\smallskip
\resizebox{\textwidth}{!}{%
\begin{tabular}{l l l l l}
\toprule
Device class & Interface challenges & Representative modeling strategies & Key clinical outputs & References \\
\midrule
FD & 
\begin{tabular}[t]{@{}l@{}}Wall apposition, foreshortening, \\ branch jailing, neck-plane MCR\end{tabular} & 
\begin{tabular}[t]{@{}l@{}}Fast virtual deployment, parametric braids, \\ beam-resolved mechanics, porous/resolved CFD\end{tabular} & 
\begin{tabular}[t]{@{}l@{}}MCR, pore density, \\ wall shear stress (WSS), inflow reduction\end{tabular} & 
\cite{ma2012computer,kim2024quantitative} \\
\addlinespace
WEB & 
\begin{tabular}[t]{@{}l@{}}Ostium coverage, residual neck filling, \\ device sizing, chronic shape changes\end{tabular} & 
\begin{tabular}[t]{@{}l@{}}Image-based sizing, geometric placement, \\ device-resolved/porous CFD, deformation tracking\end{tabular} & 
\begin{tabular}[t]{@{}l@{}}Covered neck area, residual filling, \\ kinetic energy reduction, deformation rate\end{tabular} & 
\cite{shah2021volume,munoz2024modification} \\
\addlinespace
CNS & 
\begin{tabular}[t]{@{}l@{}}Neck-rim seating, exposed ostium, \\ deformation, migration, recurrence\end{tabular} & 
\begin{tabular}[t]{@{}l@{}}Fast placement, contact-resolved deployment, \\ image-based follow-up, coverage-dependent CFD\end{tabular} & 
\begin{tabular}[t]{@{}l@{}}Uncovered ostium area, inflow reduction, \\ migration metrics, deformation indices\end{tabular} & 
\cite{lyu2024treatment,pramanik2026contact} \\
\addlinespace
Emerging & 
\begin{tabular}[t]{@{}l@{}}Novel geometry optimization, \\ apposition stability, flow disruption\end{tabular} & 
\begin{tabular}[t]{@{}l@{}}Bench deployment, parametric design loops, \\ high-throughput CFD screening, structural validation\end{tabular} & 
\begin{tabular}[t]{@{}l@{}}Deployment feasibility, velocity reduction, \\ positional sensitivity maps\end{tabular} & 
\cite{hecker2025artisse,zoppo2024novel,Voss2025} \\
\bottomrule
\end{tabular}%
}
\end{table*}

\section{Mechanics of the deployed implant state}
\label{sec:mechanics_deployment}

Virtual deployment establishes the boundary condition for every subsequent flow calculation. Depending on the question, a model may need to represent radial expansion, inter-wire slip, contact with the vessel or aneurysm wall, and the pore geometry across the neck. Fast approaches prescribe or geometrically adapt a device configuration. Higher-fidelity approaches resolve braid kinematics, beam mechanics, frictional contact, superelastic material response, and parts of the procedural history. Neither level is universally preferable; the useful level is the least complex model that still resolves the mechanism under study.

\subsection{Braid mechanics}
\label{subsec:braided_beams}

\begin{figure}[htpb!]
\centering
\includegraphics[width=0.9\textwidth]{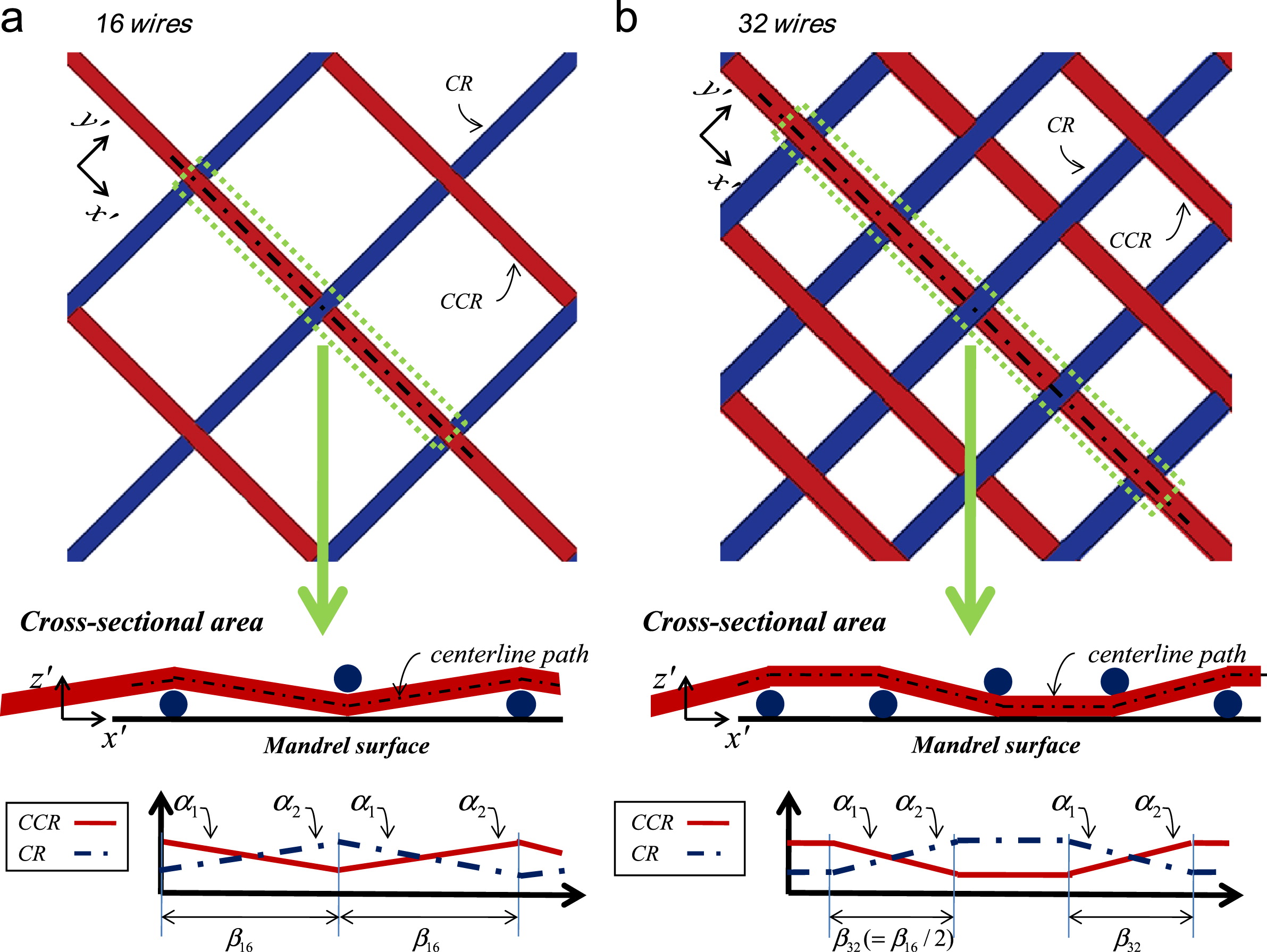}
\caption{Wire-scale architecture of braided intraluminal FD: simplified and higher-density braid representations; the local braid angle $\alpha$ changes during expansion or compression. Exact wire count increases the number of beam elements and crossover interactions, whereas a reduced braid could be sufficient for early design studies. Reproduced with permission from~\cite{kim2008mechanical}.}
\label{fig:braid_architectures}
\end{figure}

Braided neurovascular implants contain repeated interwoven wire paths and a large number of crossovers (Figs.~\ref{fig:beds_braids} and~\ref{fig:braid_architectures}). A three-dimensional solid model of every wire requires several elements across each small wire diameter and contact constraints at each crossover, which makes repeated patient-specific analyses expensive. Geometrically exact beam theory offers a practical alternative: each wire is represented by its centerline while retaining axial stretch, shear, bending, torsion, finite rotation, and contact with neighboring wires and surfaces~\cite{antman2005nonlinear,simo1986finite,jelenic1999objectivity,meier2015locking,meier2014objective,meier2019geometrically,frank2026mechanical,konyukhov2018consistent,lalonde2017modeling,bali2025finite,ait2025three}.

Exact device-specific wire count increases the number of beam elements, crossover interactions, and contact conditions. A reduced braid can accelerate early design studies, but it also changes stiffness and pore topology. Model resolution should therefore be selected against the mechanical or clinical quantity to be predicted rather than against geometric detail alone.

Beam theory itself introduces choices. A shear-deformable formulation is robust for general curved-wire configurations, whereas a slender Kirchhoff-type beam can reduce degrees of freedom when shear deformation is negligible. Element order and centerline resolution must be sufficient to capture curvature without producing artificial local bending at the crossover scale. These decisions affect not only convergence but also the contact geometry, because the distance and orientation between neighboring centerlines determine when contact is detected.

Neurovascular braids range from FDs with approximately 48--96 wires to intrasaccular devices with as many as 144 wires \cite{gaub2024flow,bhogal2019endosaccular}. Their simulation combines finite-deformation beam mechanics, multi-body contact, and coupling between one-dimensional wires and higher-dimensional anatomical surfaces or volumes. Geometrically exact Cosserat and Simo-Reissner beam formulations provide the mathematical basis for this treatment~\cite{steinbrecher2020,steinbrecher2022,steinbrecher2025consistent,Khristenko2021,Portillo2026}, which have also been successfully extended to mixed-dimensional coronary angioplasty environments~\cite{datz2025patient}. Open-source toolchains now make these methods more accessible~\cite{4C,BeamMe,bisighini2022endobeams}. Internal wire-wire contact determines slip and compaction, whereas wire-wall contact determines apposition, local MCR, and the neck-plane pore map passed to CFD~\cite{pramanik2026contact}.

\subsection{Crossover and contact mechanics}
\label{subsec:crossover_modeling}

The treatment of wire intersections strongly affects the simulated braid response. In a joint-crossover model, intersecting nodes are kinematically coupled~\cite{steinbrecher2026variationally}. This is efficient for design sweeps but suppresses relative sliding and can bias the structure toward a stiffer response. Woven-crossover models instead resolve intersections through contact, often with friction. They are more nonlinear and parameter-sensitive, but can represent local slip, rearrangement, compaction, and inversion~\cite{kelly2019comparison,shanahan2017looped}.

The insertion sequence adds a second layer of path dependence. Catheter crimping, navigation through tortuous vessels, push-pull adjustments, and sheath retraction alter the device stress state before final release~\cite{ma2012computer}. Curvature and size mismatch combine bending, torsion, and axial deformation. Smooth centerline descriptions based on isogeometric analysis and non-uniform rational B-splines are useful for tracking through highly curved anatomy~\cite{chavalla2019simulation,do2024numerical}. Fast kinematic deployment avoids much of this procedural history and remains valuable for screening, provided that its geometric predictions are validated against bench experiments or post-operative imaging before patient-specific interpretation.

\begin{figure}[htpb!]
\centering
\includegraphics[width=\textwidth]{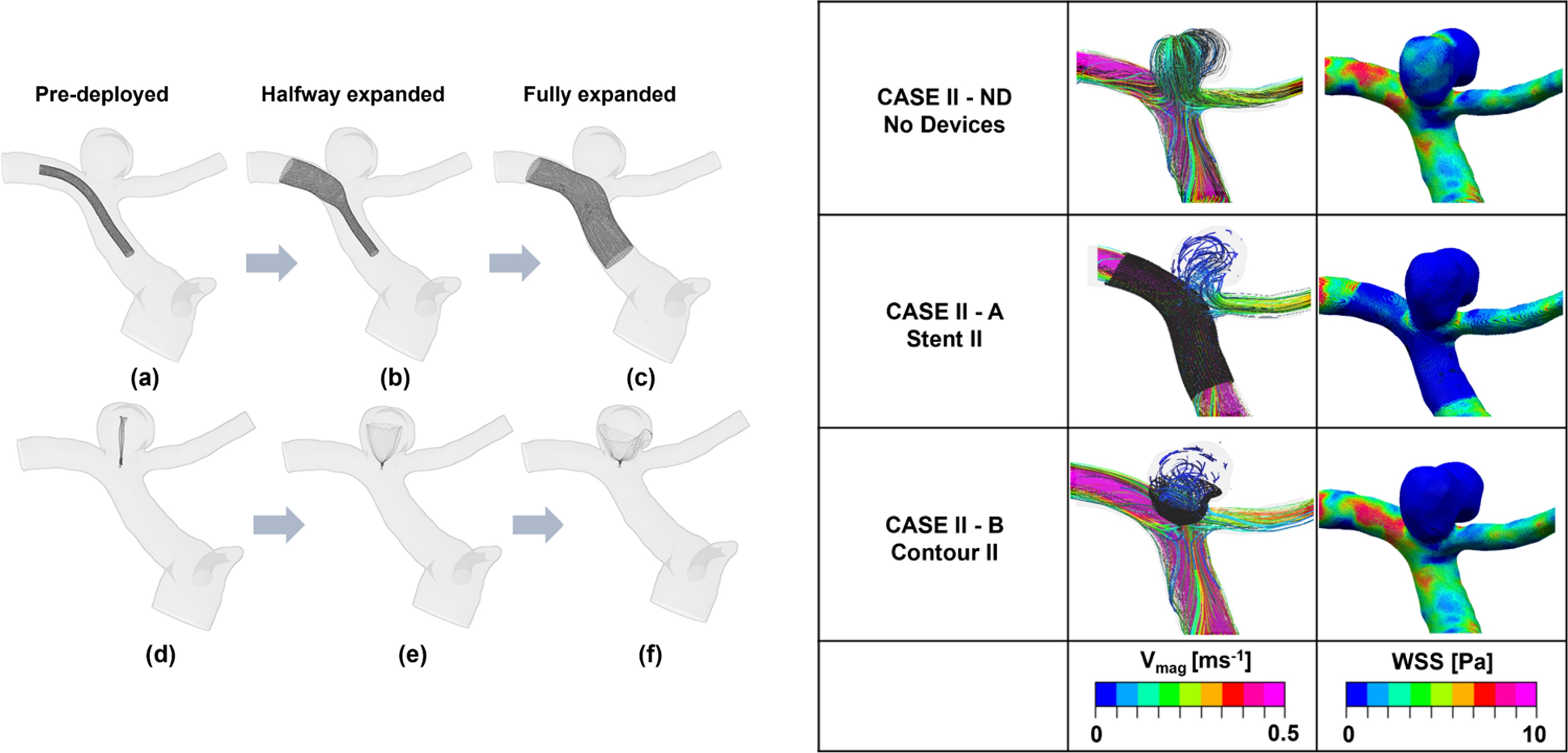}
\caption{Kinematic virtual deployment and corresponding hemodynamic comparison between an intraluminal FD and an intrasaccular CNS device. The left panel group shows staged deployment, and the right panel group contrasts the untreated aneurysm (top), the intraluminal FD (middle), and the intrasaccular CNS device (bottom) using velocity-magnitude and WSS fields. Both treatments attenuate the intra-aneurysmal jet but differ in their effects on the branching vessels. Reproduced with permission from~\cite{lyu2024treatment}.}
\label{fig:fd_cns_fast_deployment}
\end{figure}

Figure~\ref{fig:fd_cns_fast_deployment} provides a concrete example of such a fast kinematic workflow. Staged placement of an intraluminal FD and an intrasaccular CNS device is paired with the corresponding velocity and WSS fields, making clear both the efficiency of the approach and its dependence on idealized positioning and boundary assumptions.

Contact resolution is usually the most expensive and least easily calibrated part of this pathway. Penalty parameters, contact search tolerances, effective wire radius, and friction all influence convergence and local slip. A numerically stable deployment is not necessarily a mechanically credible one: excessive penalty stiffness can lock crossovers, whereas a contact tolerance that is too large can create premature interaction. Reporting these algorithmic choices, together with a friction sensitivity study, is therefore necessary when local compaction or anchoring is part of the conclusion~\cite{kelly2019comparison,alherz2016numerical}.

Parametric and analytical formulations complement these continuum solvers. Closed-form helical equations, geometric optimization, and spring-based stiffness models accelerate early design iterations and sensitivity analyses~\cite{de2009virtual,zhang2016towards,abdollahi2024virtual,suzuki2017relationships,qiu2022influence,shang2023bending,zaccaria2021analytical}. They are less suited to questions that depend on local contact, such as migration, compaction, micro-malapposition, or isolated pore-size anomalies at the ostium.

\subsection{Implant–vasculature interaction}
\label{subsec:device_vasculature_interface}

Wire-wall contact and interfacial friction govern apposition, ostium coverage, and mechanical stability. A braid may have favorable nominal porosity in an unconstrained bench configuration yet develop local malapposition after deployment, creating a preferential inflow path. Vessel diameter, curvature, axial compression, and wall contact all modify MCR and pore geometry~\cite{dholakia2017hemodynamics}. Friction affects both sliding at internal crossovers and tangential anchoring at the wall. Its coefficient should therefore be reported and, where possible, calibrated against bench data or explored through sensitivity analysis.

A single friction coefficient should not automatically be transferred across devices. Electropolishing, oxide condition, polymer coating, wire pairing, wall material, and lubrication all affect the interface measured on the bench. In a patient-specific model, friction is often better treated as an uncertain parameter than as a precisely known material constant. The useful output may then be a range of possible apposition or migration responses rather than one deterministic configuration.

For FDs, device-artery interaction controls apposition, foreshortening, and branch-vessel coverage~\cite{fu2017interaction}. For WEB and CNS devices, ostium coverage is a central interface metric because it defines the resistance presented to incoming flow. Image-based studies associate uncovered ostium area with intra-aneurysmal velocity and inflow transport~\cite{korte2025analysis,spitz2024assessment}, while longitudinal imaging shows that device shape and position can continue to change after treatment~\cite{larsen2026contour}. Explicit crossover and pore-level resolution is therefore most valuable when the question concerns compaction, slip, persistent deformation, or local coverage loss.

\subsection{Superelastic Nitinol}
\label{subsec:superelastic_materials}

Nitinol is widely used in self-expanding neurovascular implants. Above the austenite finish temperature, loading induces transformation from austenite to martensite; unloading drives the reverse transformation and enables pseudoelastic recovery~\cite{duerig1999nitinol,OTSUKA2005}. This mechanism accommodates the large recoverable strains imposed by catheter crimping, tortuous navigation, and release. Medical-grade nitinol can recover nonlinear strains of approximately 8\%, well beyond the elastic range of conventional metallic alloys~\cite{pelton2000optimization,duerig2000overview}.

Anatomical constraint can prevent complete recovery after deployment. Because a self-expanding implant is commonly oversized relative to the parent artery or aneurysm ostium, its deployed state remains below the unconstrained diameter. The resulting chronic outward force and residual martensitic fraction depend on the local boundary conditions~\cite{pelton2022prestrain}. Material state, radial force, and interface contact are therefore coupled.

Not every simulation requires a full phase-transformation law. Linear elasticity may be adequate for an early geometric study if wire strains remain small and the output is not force-dependent. It becomes difficult to justify after severe crimping or when the study predicts radial force, recovery, hysteresis, residual stress, or fatigue. The material model should therefore be chosen from the expected strain path and the intended output, not simply from the nominal device material.

Nitinol response also depends on processing. Cold working, heat treatment, grain size, alloy composition, wire drawing, laser cutting, and electropolishing alter transformation plateaus, fatigue resistance, and surface chemistry~\cite{ma16196480,HWANG1983381,Labu,GALL20024643,elsisy2020materials}. Structural fatigue concerns crack initiation and damage accumulation; functional fatigue concerns loss of recoverable strain, shifts in transformation temperature, and changes in plateau force. After deployment, cyclic loading arises from pulsatile flow and vessel-wall motion~\cite{Pelton2008}. A model intended to predict force, recovery, or durability therefore needs a constitutive law calibrated to the actual wire and its processing history, including transformation stress, hysteresis, temperature, and pre-strain where relevant~\cite{auricchio1997shape,Bernini,weafer2013micro}.

\section{From deployment to hemodynamics}
\label{sec:hemodynamics_multiphysics}

Once deployed, the implant becomes a geometrical boundary within a fluid domain. The appropriate fluid model again depends on the question. Rigid-wall CFD can provide useful relative metrics for device screening and controlled comparisons. With sufficient geometric resolution, it can also capture pore-scale flow, malapposition jets, and pulsatile patterns. FSI becomes relevant when wall compliance, tissue stress, device motion, or reciprocal fluid-solid loading forms part of the hypothesis.

\subsection{Rigid-wall CFD}
\label{subsec:rigid_wall_cfd}

Rigid-wall CFD (i.e., numerical simulations of the flow field in intracranial aneurysms assuming the vessel walls to be rigid) remains the dominant approach because it is robust and comparatively inexpensive~\cite{berg2019review,kono2013hemodynamics}. It solves the Navier-Stokes equations in a static lumen and is well suited to comparisons of velocity, inflow, WSS, oscillatory shear index (OSI), and transport under controlled conditions. The limitation is explicit: the wall does not move. This suppresses pressure-diameter coupling and may alter local flow indicators~\cite{takizawa2012comparative,torii2009fluid,eken2017parallel}. CFD also cannot provide tissue stress, so it is not the appropriate model when mural stress is the endpoint.

Velocity and shear are solved on a predetermined geometry, so their interpretation remains sensitive to the structural assumptions that created that geometry. A refined fluid solver cannot recover deployment information that was never represented. For questions involving local coverage or malapposition, studies should therefore quantify whether an interface-resolved deployment changes the conclusion relative to a simpler placement method.

\begin{figure}[htpb!]
\centering
\includegraphics[width=\textwidth]{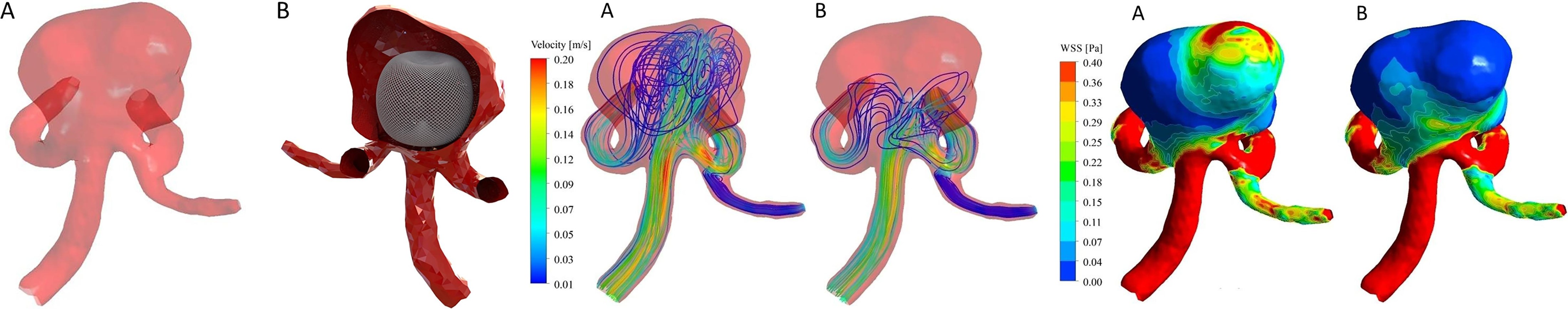}
\caption{Image-based CFD workflow for an anterior communicating artery aneurysm treated with an intrasaccular WEB device. Within each image pair, \textbf{A} denotes the untreated or pre-deployment model and \textbf{B} denotes the WEB-treated or post-deployment model. From top to bottom, the rows show the reconstructed vascular geometry, velocity streamlines, and WSS distributions, illustrating how the implanted mesh changes intra-aneurysmal flow and near-wall shear. Reproduced with permission from~\cite{aghli2021image}.}
\label{fig:web_cfd_workflow}
\end{figure}

An image-based WEB example makes this geometric dependency explicit (Fig.~\ref{fig:web_cfd_workflow}). Even when the vascular reconstruction and fluid solver are otherwise unchanged, insertion of the deployed mesh modifies the intra-aneurysmal streamlines and redistributes near-wall shear. The comparison is therefore interpretable only together with the assumed post-deployment device configuration.

Boundary conditions remain another major source of variation. Patient-specific inlet waveforms acquired with flow-sensitive Magnetic Resonance Imaging (MRI) methods are preferable when available, but they do not remove uncertainty in distal resistance or pressure. Generic waveforms can still support controlled device comparisons if the limitation is stated and identical conditions are used across cases. The more problematic practice is to interpret absolute WSS or residence time without reporting inlet scaling, outlet treatment, cardiac-cycle convergence, and the sensitivity of the conclusion to those choices~\cite{berg2019review}.

\begin{figure}[htpb!]
\centering
  \begin{subfigure}{\textwidth}
    \centering
    \includegraphics[width=\textwidth]{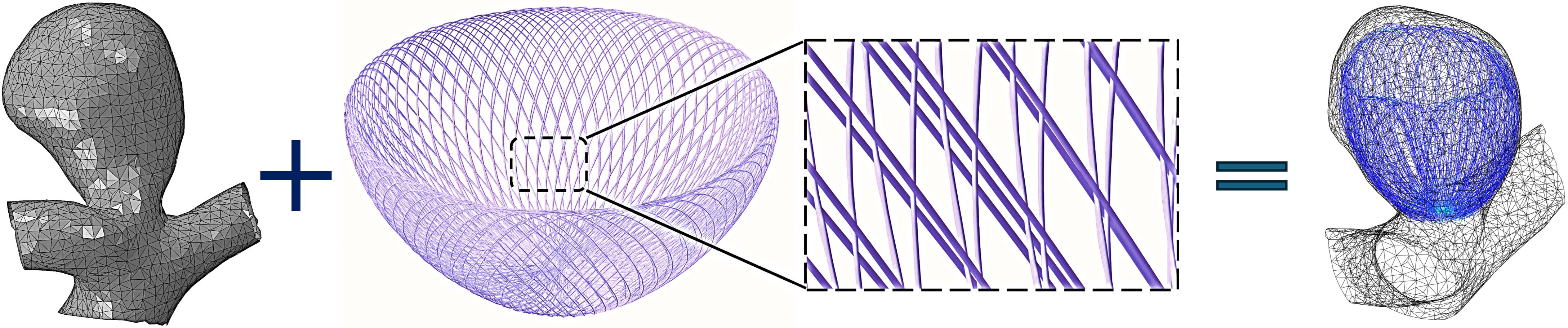}
    \caption{\textit{In silico} placement of a CNS device within a patient-specific aneurysm model.}
    \label{fig:my_cns}
  \end{subfigure}
  \vspace{1em}
  \begin{subfigure}{0.48\textwidth}
    \centering
    \includegraphics[width=\textwidth]{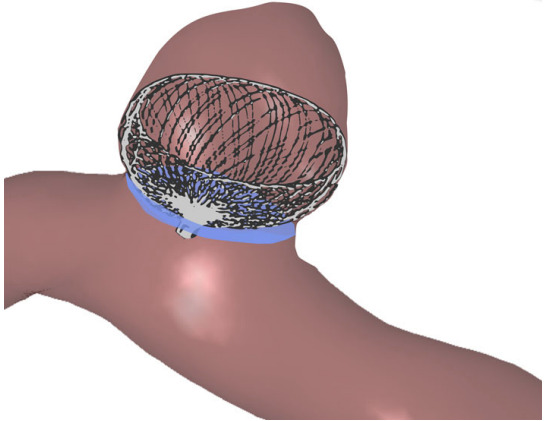}
    \caption{Fast geometric placement of the implant.}
    \label{fig:bottom_fast_placement}
  \end{subfigure}\hfill
  \begin{subfigure}{0.48\textwidth}
    \centering
    \includegraphics[width=\textwidth]{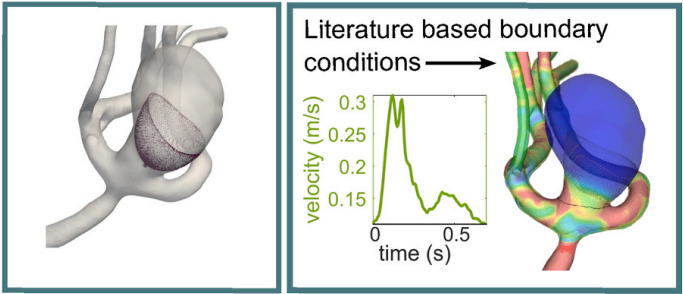}
    \caption{Post-treatment CFD simulation velocity fields.}
    \label{fig:bottom_hemodynamics}
  \end{subfigure}
\vfill
\caption{Propagation of deployment geometry into downstream hemodynamic analysis. (a) End-to-end mechanics-to-flow pipeline linking virtual implant placement and structural wire-resolved meshes to post-treatment flow evaluation. (b, c) Patient-specific geometric configuration of a CNS, establishing the localized ostium coverage and residual structural boundaries that directly govern the subsequent CFD velocity profiles, reproduced from previous work by our co-author~\cite{spitz2024assessment,korte2025analysis}.}
\label{fig:cnt_fast_placement_hemodynamics}
\end{figure}

\subsection{Device representation}
\label{subsec:device_representation_scale}

Post-treatment hemodynamics span very different length scales: individual wires are approximately $15$--$50\,\upmu\mathrm{m}$ in diameter, whereas the parent vessel is typically $2$--$5\,\mathrm{mm}$ wide. A device-resolved calculation therefore needs fine elements near each wire. Porous-medium and heterogeneous-domain models reduce this cost by replacing explicit wires with a local resistance or permeability tensor~\cite{abdehkakha2021cerebral,berod2022heterogeneous}. These approximations are useful for cohort screening and global flow reduction. Explicit wires are needed when the endpoint is near-wire flow, a local malapposition gap, or a high-velocity boundary jet~\cite{reymond2025novel}.

Figure~\ref{fig:cnt_fast_placement_hemodynamics} shows how device placement propagates into the flow calculation. Local resistance depends nonlinearly on wire density and pore size, and errors in uncovered ostium area can therefore produce disproportionate changes in simulated inflow. Advanced solvers, including lattice-Boltzmann methods, can resolve fine flow structures but still inherit uncertainty from the post-deployment geometry and its wall apposition~\cite{frank2024numerical}.

Patient-specific FSI that includes slender implant wires remains uncommon. Mixed-dimensional methods provide a route forward by coupling one-dimensional beams to a three-dimensional incompressible flow domain. Current formulations range from one-way coupling to strongly coupled partitioned schemes with regularized interface transfer~\cite{hagmeyer2022oneway,hagmeyer2024fullycoupled,lespagnol2024mixed}. Their value will depend on showing when fluid-solid feedback changes a clinically relevant conclusion rather than only increasing numerical fidelity.

One-way coupling can be sufficient when fluid forces are too small to appreciably alter the device or wall configuration over the simulated period. Two-way coupling becomes more relevant when wall motion changes the lumen, when an implant moves relative to the tissue, or when the endpoint is an exchanged load. This distinction should be made before selecting a coupling algorithm. Otherwise, numerical effort can increase substantially without improving the clinical interpretation.

\subsection{Wall mechanics, blood rheology, and thrombosis}
\label{subsec:wall_rheology_thrombus}

Wall compliance is one reason to move beyond a rigid domain. An FSI study of 101 patient-specific aneurysms found differences between rigid- and deformable-wall models in near-dome recirculation and sac-averaged OSI~\cite{goetz2024anxplore}. Other patient-specific studies likewise show changes in vortex structure and in the magnitude of hemodynamic metrics when wall mechanics are included~\cite{shidhore2023comparative,jeken2025investigating}. These findings support a conditional choice: rigid-wall CFD remains useful for standardized comparison, whereas FSI is appropriate when wall motion, tissue stress, or fluid-solid loading is central to the question.

Blood rheology introduces further uncertainty. A Newtonian model is often adequate in the higher-shear parent vessel, whereas shear-thinning may become more relevant inside a treated sac as velocity falls and residence time increases~\cite{campo2015review,saqr2020computational,boniforti2024endovascular}. Thrombus formation is harder still. Most workflows use early reductions in velocity, inflow, or shear as physical surrogates for occlusion and do not resolve platelet activation, coagulation, thrombus remodeling, or endothelial healing~\cite{ngoepe2018thrombosis}. Multiphase and scalar-transport models begin to address this gap, but clinical validation remains limited~\cite{lampropoulos2025investigating,chowdhury2025flow}.

This distinction between a surrogate and an outcome is clinically important. A large immediate reduction in inflow may favour thrombosis, but it does not by itself predict durable occlusion. Medication, surface properties, aneurysm size, thrombus organization, and tissue response remain outside most flow models. Hemodynamic results should therefore be framed as mechanistic evidence about the acute environment unless they have been linked to longitudinal outcomes in an appropriately validated cohort.

Post-deployment hemodynamic quantities should therefore be presented as model-dependent biophysical indicators. Velocity reduction, WSS, oscillatory shear, inflow, and residence time remain interpretable only when the deployed geometry, inlet and outlet conditions, wall model, rheology, and device representation are reported alongside them.

\section{From computational models to clinical prediction}
\label{sec:discussion}

\subsection{Current limitations}
\label{subsec:limitations}

An important asymmetry remains in many endovascular simulations: the flow field may be resolved with millions of elements, while the implant geometry is prescribed with comparatively little mechanical information. This is not always a problem. A simplified geometry may answer a comparative screening question well. It becomes a problem when the conclusion depends on local apposition, pore size, compaction, or contact, because these features define the fluid boundary itself.

For FDs, commonly simplified quantities include procedural history, wall anchoring, non-uniform foreshortening, and local changes in MCR. For WEB and CNS devices, models often approximate device seating at the ostium, neck-rim coverage, and delayed deformation. Emerging implants face a different limitation: computational screening and bench expansion can establish feasibility, but they do not by themselves establish durability at the device-wall interface.

Geometry is therefore a source of model uncertainty, not a cosmetic preprocessing step. Device-sizing studies show that MCR and pore-density distributions are sensitive to mismatch between device and parent-vessel diameter~\cite{kim2024quantitative}. Micro-computed tomography~\cite{reymond2025novel} and optical measurements of \textit{in vitro} deployment reveal irregular, non-periodic braid configurations that differ from idealized virtual placements~\cite{velvaluri2021realistic}. The practical question is not whether every study needs full deployment mechanics, but whether uncertainty in the deployed geometry is large enough to change the intended interpretation.

\subsection{Interface durability}
\label{subsec:durability_interface}

For intrasaccular implants, the device-ostium interface evolves after treatment. Initial coverage reduces inflow and creates conditions for thrombus organization and tissue healing. Later deformation or displacement can reduce that coverage and reopen preferential inflow paths (Table~\ref{tab:durability_loop_simple}). Longitudinal studies of WEB and CNS devices show changes in shape and position over time, and associate deformation or migration with recurrence~\cite{munoz2024modification,griessenauer2025contour,wodarg2025embolization}.

\begin{table*}[htpb!]
\centering
\caption{Time-dependent interface between device mechanics, flow, and treatment durability.}
\label{tab:durability_loop_simple}
\smallskip
\resizebox{\textwidth}{!}{%
\begin{tabular}{l l l l}
\toprule
Process & Core elements & Biomedical interpretation & References \\
\midrule
Initial interface &
\begin{tabular}[t]{@{}l@{}}Ostium occlusion matrix, device sizing, \\ contact and friction dictating initial coverage\end{tabular} &
\begin{tabular}[t]{@{}l@{}}Establishes the starting boundary for acute \\ flow attenuation and device stability\end{tabular} &
\cite{korte2025analysis,larsen2026contour} \\
\addlinespace
Acute flow response &
\begin{tabular}[t]{@{}l@{}}Inflow attenuation, WSS and OSI shifts, \\ sensitivity to boundary and wall assumptions\end{tabular} &
\begin{tabular}[t]{@{}l@{}}Links neck coverage to early hemodynamic \\ surrogates of occlusion\end{tabular} &
\cite{dholakia2017hemodynamics,spitz2024assessment} \\
\addlinespace
Remodeling &
\begin{tabular}[t]{@{}l@{}}Thrombus maturation, tissue healing, \\ persistent pulsatile cyclic mechanical loading\end{tabular} &
\begin{tabular}[t]{@{}l@{}}Converts the acute interface into a remodeling, \\ time-dependent biomechanical system\end{tabular} &
\cite{ngoepe2018thrombosis,pelton2022prestrain} \\
\addlinespace
Coverage loss &
\begin{tabular}[t]{@{}l@{}}Interfacial coverage loss over time, \\ recurrence risk via late-stage inflow reopening\end{tabular} &
\begin{tabular}[t]{@{}l@{}}Explains how delayed deformation, compaction, \\ or migration can reverse early treatment benefit\end{tabular} &
\cite{griessenauer2025contour,wodarg2025embolization} \\
\bottomrule
\end{tabular}%
}
\end{table*}

The complete inversion of a CNS device observed during follow-up illustrates why the post-deployment state cannot always be treated as permanent~\cite{bellanger2024reversed}. The mechanism of slower changes remains less certain. Pulsatile loading, wall motion, inter-wire slip, evolving thrombus, and tissue healing may all contribute. A realistic near-term goal is therefore to test the sensitivity of long-term stability to contact, friction, material state, and wall constraint, and to compare the predicted direction and magnitude of change with longitudinal imaging. Multi-year fatigue or remodeling predictions should be viewed as prospective until such validation becomes available.

\subsection{Validation, uncertainty, and reporting}
\label{subsec:validation_reporting}

Validation should focus on measurable quantities that relate to the model's intended use. Neck coverage and pore structure can be measured with optical imaging, micro-computed tomography, or by superimposing digital models onto physical phantom images. Radial force, compression, and hysteresis can be obtained from bench tests~\cite{dengiz2025thin}. Particle image velocimetry, DSA-derived transport measures, and four-dimensional flow magnetic resonance imaging are used to experimentally assess flow within an aneurysm~\cite {sindeev2018phasecontrast,Hadad2023,korte2024vitro,pravdivtseva20213d,bisighini2022fabrication}. Table~\ref{tab:validation_reporting} combines these targets with a minimum reporting framework.

No single experiment validates the complete workflow. A useful hierarchy begins with unit tests of material and contact behavior, proceeds to deployment in idealized and patient-specific phantoms, and then evaluates flow under matched geometry and boundary conditions. Clinical imaging provides the final test of position, apposition, deformation, or migration, but usually at lower spatial resolution than bench imaging. Agreement at one level should not be taken as validation of another: matching global radial force does not guarantee local pore geometry, and matching velocity in a rigid phantom does not validate a deformable-wall FSI model.

Axial compression, bending, and segmental radial-force tests are particularly useful for calibrating structural models~\cite{kim2008mechanical,kutbay2023quantitative}. Work on peripheral and coronary braided stents also provides transferable evidence on crossover slip, coatings, composite wires, radial support, plaque-driven mechanical states~\cite{datz2026influence}, and flexural behavior~\cite{giuliodori2021numerical,mckenna2021finite,hu2024modified,ubachs2023computational,zhao2024finite,zhao2025mechanical}. Wire diameter, filament count, braid angle, coating compliance, end constraints, and surface topography can then be related to apposition, migration resistance, trackability, and geometric stability.

\begin{table*}[htpb!]
\centering
\caption{Minimum reporting and validation framework for interface-resolved modeling.}
\label{tab:validation_reporting}
\smallskip
\resizebox{\textwidth}{!}{%
\begin{tabular}{l l l l}
\toprule
Model component & Minimum reporting & Validation target & References \\
\midrule
Device architecture &
\begin{tabular}[t]{@{}l@{}}Device class and size, wire count and diameter, \\ braid angle, end constraints, coatings\end{tabular} &
\begin{tabular}[t]{@{}l@{}}Optical imaging or $\upmu$CT of wire paths, \\ pore map, MCR, and neck coverage\end{tabular} &
\cite{kim2008mechanical,velvaluri2021realistic,reymond2025novel} \\
\addlinespace
Deployment and contact &
\begin{tabular}[t]{@{}l@{}}Crimping and release protocol, delivery path, size mismatch, \\ joint or woven crossover, wall contact, friction\end{tabular} &
\begin{tabular}[t]{@{}l@{}}Bench deployment and post-operative imaging of \\ position, apposition, foreshortening, and deformation\end{tabular} &
\cite{kelly2019comparison,do2024numerical,fu2017interaction} \\
\addlinespace
Material response &
\begin{tabular}[t]{@{}l@{}}Constitutive law and parameters, reference temperature, \\ transformation thresholds, hysteresis, pre-strain\end{tabular} &
\begin{tabular}[t]{@{}l@{}}Force-diameter curves, loading-unloading hysteresis, \\ recovery and cyclic response of the actual wire\end{tabular} &
\cite{auricchio1997shape,Bernini,pelton2022prestrain,kutbay2023quantitative} \\
\addlinespace
Fluid model and boundaries &
\begin{tabular}[t]{@{}l@{}}Resolved or porous device, rigid or deformable wall, rheology, \\ inlet and outlet conditions, grid and cycle convergence\end{tabular} &
\begin{tabular}[t]{@{}l@{}}Particle image velocimetry, 4D flow MRI, DSA-derived \\ transport, and comparisons across solvers\end{tabular} &
\cite{berg2019review,sindeev2018phasecontrast,korte2024vitro,goetz2024anxplore} \\
\addlinespace
Uncertainty and interpretation &
\begin{tabular}[t]{@{}l@{}}Segmentation tolerance, parameter bounds, sensitivity results, \\ and distinction between indicator and outcome prediction\end{tabular} &
\begin{tabular}[t]{@{}l@{}}Repeat analysis across anatomies and users; compare shape, \\ migration, deformation, and recurrence longitudinally\end{tabular} &
\cite{Voss2019,Paritala2023,munoz2024modification,larsen2026contour} \\
\bottomrule
\end{tabular}%
}
\end{table*}

Rigid vascular phantoms suit rigid-wall CFD validation, whereas compliant tissue-mimicking phantoms are needed when an FSI model claims to reproduce pressure-diameter coupling or energy dissipation~\cite{Yalman2025,Karam2023,Zimmermann2021}. Validation should also include the preprocessing workflow. Segmentation, smoothing, and branch truncation vary between users and can shift velocity, ostium inflow, and vortex structure even when the numerical solver is unchanged~\cite{Voss2019,Paritala2023}.

Uncertainty should be propagated in the same order as the workflow. Anatomical uncertainty changes contact; contact uncertainty changes the deployed pores; and pore uncertainty changes flow. Reporting only mesh convergence at the final CFD stage therefore gives an incomplete picture. For clinical translation, studies should identify which upstream uncertainty can reverse a device ranking or cross a clinically meaningful threshold. This decision-focused view is more informative than attaching the same exhaustive sensitivity analysis to every model component.

\section{Outlook}
\label{subsec:future_directions}

\begin{figure}[htpb!]
\centering
\includegraphics[width=0.6\textwidth]{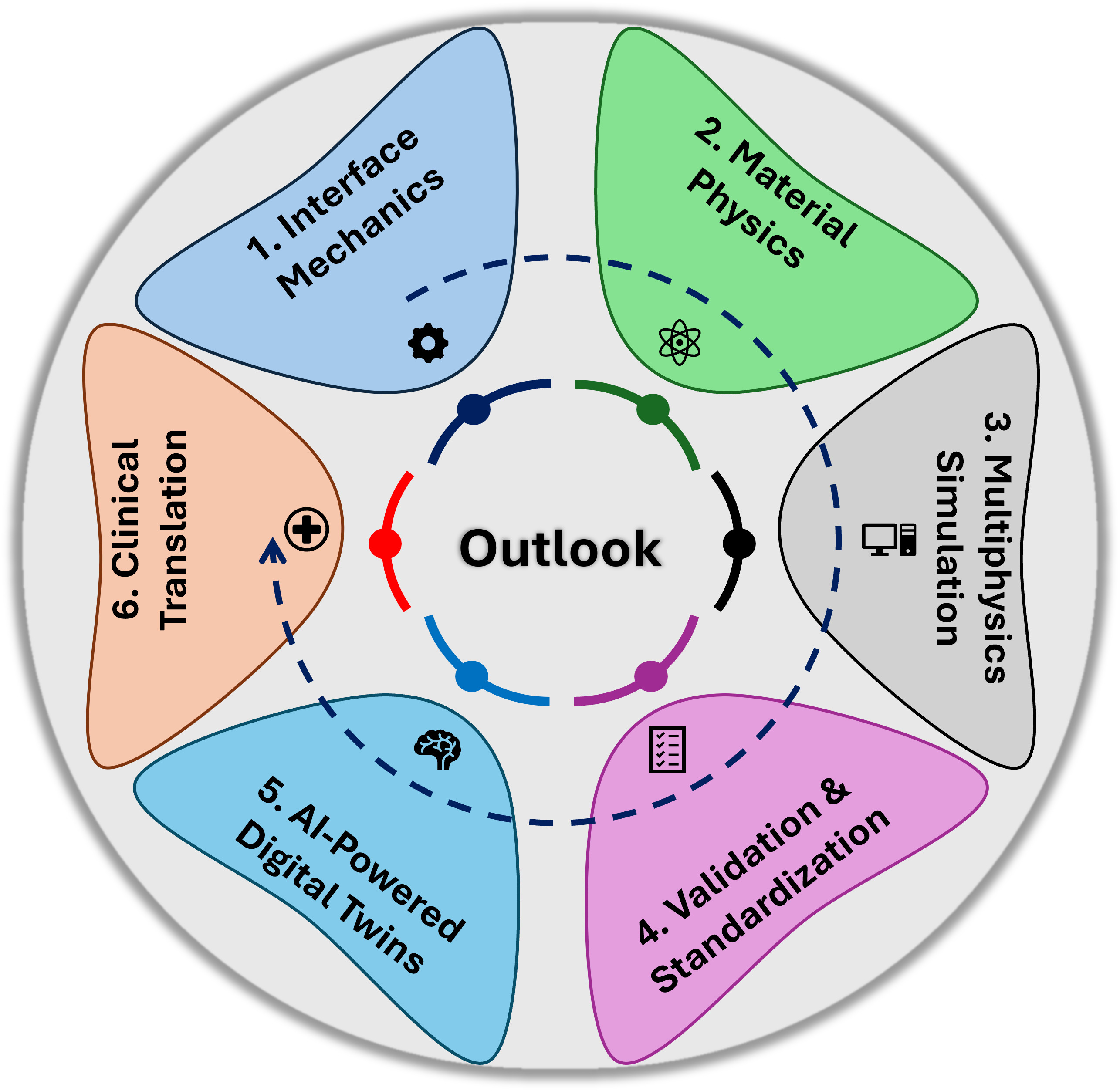}
\caption{Roadmap for interface-resolved modeling of braided endovascular implants. Six linked priorities support clinical translation: interface mechanics, material physics, multiphysics simulation, validation and standardization, AI-powered digital twins, and clinical translation.}
\label{fig:outlook}
\end{figure}

The next generation of patient-specific workflows should be multiscale, question-driven, and validation-led (see Fig.~\ref{fig:outlook}). The six priorities represented in the figure can move the field toward clinical use:
\begin{itemize}
  \item \textbf{Interface mechanics:} Generate post-operative device-vessel configurations through contact-resolved deployment when neck coverage, wall apposition, compaction, or migration is the endpoint.

  \item \textbf{Material physics:} Incorporate crimping pre-strain, deployment mean strain, hysteresis, and residual martensitic fraction when radial force, deformation, or durability is being predicted.

  \item \textbf{Multiphysics simulation:} Progress from porous screening models to device-resolved CFD or FSI only when the additional physics addresses the clinical question; extend these models to thrombosis, healing, or inflammatory mechanisms when such biology is the endpoint.

  \item \textbf{Validation and standardization:} Establish shared reporting requirements and experimental, imaging, and clinical validation targets for device geometry, apposition, flow reduction, deformation, migration, and longitudinal occlusion.

  \item \textbf{AI-powered digital twins:} Combine mechanically plausible deployment, multiphysics simulation, longitudinal imaging, and patient-specific clinical data into computational surrogates that quantify uncertainty and enable rapid comparison of treatment options.

  \item \textbf{Clinical translation:} Evaluate several plausible devices and sizes rather than reporting a single post hoc simulation, and determine whether model predictions improve treatment selection and patient outcomes.
\end{itemize}

The required building blocks already exist, but they have not yet been assembled and validated as an end-to-end clinical workflow. A pragmatic route is to use fast placement or surrogate models for initial screening, benchmark them against contact-resolved mechanics in representative anatomies, and carry the resulting geometric uncertainty into the hemodynamic interpretation~\cite{bisighini2023machine,bisighini2023patient}. This hierarchy is more realistic than applying the most expensive model to every case.

A clinically usable system would also have to operate within the time constraints of treatment planning. It should return not only a preferred device but the reason for that preference, the uncertainty in the ranking, and the failure mode that remains most plausible. Such a system is unlikely to emerge from CFD, structural mechanics, or machine learning alone. It will require prospective comparison with deployment imaging, bench measurements that reflect procedural use, and outcome data collected in a form that can be related to the modeled interface.

Across FDs, WEB, CNS, and emerging flow disruptors, post-deployment geometry reflects device architecture, superelastic recovery, procedural history, inter-wire interaction, wall contact, and friction. These factors shape coverage, pore topology, apposition, flow attenuation, and stability. The central translational principle is therefore simple: model fidelity should follow the clinical question. Contact-resolved deployment is most valuable for coverage, apposition, compaction, and migration; rigid-wall CFD remains useful for controlled flow comparisons; and FSI is warranted when wall or device motion is central. Progress toward patient-specific device selection will depend less on adding complexity in isolation than on validating each level against measurable anatomical, mechanical, hemodynamic, and longitudinal targets.

\setstretch{1.0}
\bibliography{sn-bibliography}

@article{etminan2016unruptured,
  title={Unruptured intracranial aneurysms: development, rupture and preventive management},
  author={Etminan, Nima and Rinkel, Gabriel J},
  journal={Nature Reviews Neurology},
  volume={12},
  number={12},
  pages={699--713},
  year={2016},
  publisher={Nature Publishing Group UK London}
}

@article{backes2016patient,
  title={Patient-and aneurysm-specific risk factors for intracranial aneurysm growth: a systematic review and meta-analysis},
  author={Backes, Daan and Rinkel, Gabriel JE and Laban, Kamil G and Algra, Ale and Vergouwen, Mervyn DI},
  journal={Stroke},
  volume={47},
  number={4},
  pages={951--957},
  year={2016},
  publisher={Lippincott Williams \& Wilkins Hagerstown, MD}
}

@article{texakalidis2019aneurysm,
  title={Aneurysm formation, growth, and rupture: the biology and physics of cerebral aneurysms},
  author={Texakalidis, Pavlos and Sweid, Ahmad and Mouchtouris, Nikolaos and Peterson, Eric C and Sioka, Chrissa and Rangel-Castilla, Leonardo and Reavey-Cantwell, John and Jabbour, Pascal},
  journal={World Neurosurgery},
  volume={130},
  pages={277--284},
  year={2019},
  publisher={Elsevier}
}

@article{brisman2006cerebral,
  title={Cerebral aneurysms},
  author={Brisman, Jonathan L and Song, Joon K and Newell, David W},
  journal={New England Journal of Medicine},
  volume={355},
  number={9},
  pages={928--939},
  year={2006},
  publisher={Mass Medical Soc}
}

@article{goubergrits2014hemodynamic,
  title={Hemodynamic impact of cerebral aneurysm endovascular treatment devices: coils and flow diverters},
  author={Goubergrits, Leonid and Schaller, Jens and Kertzscher, Ulrich and Woelken, Thies and Ringelstein, Moritz and Spuler, Andreas},
  journal={Expert Review of Medical Devices},
  volume={11},
  number={4},
  pages={361--373},
  year={2014},
  publisher={Taylor \& Francis}
}

@article{pierot2013endovascular,
  title={Endovascular treatment of intracranial aneurysms: current status},
  author={Pierot, Laurent and Wakhloo, Ajay K},
  journal={Stroke},
  volume={44},
  number={7},
  pages={2046--2054},
  year={2013},
  publisher={Lippincott Williams \& Wilkins Hagerstown, MD}
}

@article{lauzier2023review,
  title={A review of technological innovations leading to modern endovascular brain aneurysm treatment},
  author={Lauzier, David C and Huguenard, Anna L and Srienc, Anja I and Cler, Samuel J and Osbun, Joshua W and Chatterjee, Arindam R and Vellimana, Ananth K and Kansagra, Akash P and Derdeyn, Colin P and Cross, Dewitte T and others},
  journal={Frontiers in Neurology},
  volume={14},
  pages={1156887},
  year={2023},
  publisher={Frontiers Media SA}
}

@article{muskens2017woven,
  title={The Woven Endobridge device for treatment of intracranial aneurysms: a systematic review},
  author={Muskens, Ivo S and Senders, Joeky T and Dasenbrock, Hormuzdiyar H and Smith, Timothy RS and Broekman, Marike LD},
  journal={World Neurosurgery},
  volume={98},
  pages={809--817},
  year={2017},
  publisher={Elsevier}
}

@article{heiferman2024new,
  title={A new era in the treatment of wide necked bifurcation aneurysms: intrasaccular flow disruption},
  author={Heiferman, Daniel M and Goyal, Nitin and Inoa, Violiza and Nickele, Christopher M and Arthur, Adam S},
  journal={Interventional Neuroradiology},
  volume={30},
  number={1},
  pages={31--36},
  year={2024},
  publisher={SAGE Publications Sage UK: London, England}
}

@article{akhunbay2020endovascular,
  title={Endovascular treatment of wide-necked intracranial aneurysms using the novel contour neurovascular system: a single-center safety and feasibility study},
  author={Akhunbay-Fudge, Christopher Yusuf and Deniz, Kenan and Tyagi, Atul Kumar and Patankar, Tufail},
  journal={Journal of NeuroInterventional Surgery},
  volume={12},
  number={10},
  pages={987--992},
  year={2020},
  publisher={British Medical Journal Publishing Group}
}

@article{zhuo2025intrasaccular,
  title={Intrasaccular therapy in wide-neck intracranial aneurysms: a narrative review},
  author={Zhuo, Kaiquan and Wu, Hongxia and Gou, Zhongji and Zhang, Changwei},
  journal={Frontiers in Neurology},
  volume={16},
  pages={1552848},
  year={2025},
  publisher={Frontiers Media SA}
}

@article{zhou2017complications,
  title={Complications associated with the use of flow-diverting devices for cerebral aneurysms: a systematic review and meta-analysis},
  author={Zhou, Gong and Su, Meng and Yin, Yu-Lin and Li, Ming-Hua},
  journal={Neurosurgical Focus},
  volume={42},
  number={6},
  pages={E17},
  year={2017},
}

@article{velvaluri2021torsional,
  title={Torsional Characterization of Braided Flow Diverter Stents: A New Method to Evaluate Twisting Phenomenon},
  author={Velvaluri, P. and Hensler, J. and Wodarg, F. and Jansen, O. and Quandt, E.},
  journal={Clinical Neuroradiology},
  year={2021},
}

@article{hohenstatt2020branch,
  title={Branch vessel occlusion in aneurysm treatment with flow diverter stent},
  author={Hohenstatt, S. and Arrichiello, A. and Conte, G. and Craparo, G. and Caranci, F. and Angileri, A. and Levi, D. and Carrafiello, G. and Paolucci, A.},
  journal={Acta Bio Medica: Atenei Parmensis},
  volume={91},
  pages={e2020003},
  year={2020},
}

@article{muhlbenninghaus2019transient,
  title={Transient in-stent stenosis: a common finding after flow diverter implantation},
  author={M{\"u}hl-Benninghaus, Ralph and Hau{\ss}mann, Anna and Simgen, Andrea and Tomori, Tamas and Reith, Wolfgang and Yilmaz, Uta},
  journal={Journal of NeuroInterventional Surgery},
  volume={11},
  number={2},
  pages={196--199},
  year={2019},
}

@article{larsen2026contour,
  title={Morphological factors and device deformation associated with recurrence in intracranial aneurysms treated with the contour neurovascular system},
  author={Larsen, Naomi and Wodarg, Fritz and Hensler, Johannes and Mostafa, Karim and Peters, S{\"o}nke and Pravdivtseva, Mariya and Saalfeld, Sylvia and G{\"a}rtner, Friederike},
  journal={Journal of Neuroradiology},
  volume={53},
  pages={101407},
  year={2026},
  publisher={Elsevier Masson SAS},
}

@article{kallmes2007new,
  title={A New Endoluminal, Flow-Disrupting Device for Treatment of Saccular Aneurysms},
  author={Kallmes, David F. and Ding, Yong Hong and Dai, Daying and Kadirvel, Ramanathan and Lewis, Debra A. and Cloft, Harry J.},
  journal={Stroke},
  volume={38},
  number={8},
  pages={2346--2352},
  year={2007},
}

@article{dholakia2017hemodynamics,
  title={Hemodynamics of Flow Diverters},
  author={Dholakia, Ronak and Sadasivan, Chander and Fiorella, David J and Woo, Henry H and Lieber, Baruch B},
  journal={Journal of Biomechanical Engineering},
  volume={139},
  number={2},
  pages={021002},
  year={2017},
  publisher={ASME},
}

@article{kim2024quantitative,
  title={Quantitative analysis of hemodynamic changes induced by the discrepancy between the sizes of the flow diverter and parent artery},
  author={Kim, Sunghan and Yang, Hyeondong and Oh, Je Hoon and Kim, Yong Bae},
  journal={Scientific Reports},
  volume={14},
  number={1},
  pages={10653},
  year={2024},
  publisher={Nature Publishing Group UK London}
}

@article{mantilla2023woven,
  title={The Woven EndoBridge device, an effective and safe alternative endovascular treatment of intracranial aneurysm - systematic review},
  author={Mantilla, Daniel E and D Vera, Daniela and Ortiz, Andr{\'e}s F and Nicoud, Franck and Costalat, Vincent},
  journal={Interventional Neuroradiology},
  pages={15910199231201544},
  year={2023},
  publisher={SAGE Publications Sage UK: London, England}
}

@article{monteiro2022treatment,
  title={Treatment of ruptured intracranial aneurysms with the Woven EndoBridge device: a systematic review},
  author={Monteiro, Andre and Lazar, Audrey L and Waqas, Muhammad and Rai, Hamid H and Baig, Ammad A and Cortez, Gustavo M and Dossani, Rimal H and Cappuzzo, Justin M and Levy, Elad I and Siddiqui, Adnan H},
  journal={Journal of NeuroInterventional Surgery},
  volume={14},
  number={4},
  pages={366--370},
  year={2022},
  publisher={British Medical Journal Publishing Group}
}

@article{korte2025analysis,
  title={Analysis of the treatment effect of the Contour Neurovascular System in intracranial aneurysms: Larger neck coverage area is associated with longitudinal flow reduction},
  author={Korte, Jana and Gaidzik, Franziska and Spitz, Lena and Pravdivtseva, Mariya S and Behme, Daniel and Larsen, Naomi and Saalfeld, Sylvia and Berg, Philipp},
  journal={Computers in Biology and Medicine},
  volume={197},
  pages={111002},
  year={2025},
  publisher={Elsevier}
}

@article{frank2024numerical,
  title={Numerical simulation of endovascular treatment options for cerebral aneurysms},
  author={Frank, Martin and Holzberger, Fabian and Horvat, Medeea and Kirschke, Jan and Mayr, Matthias and Muhr, Markus and Nebulishvili, Natalia and Popp, Alexander and Schwarting, Julian and Wohlmuth, Barbara},
  journal={GAMM-Mitteilungen},
  volume={47},
  number={3},
  pages={e202370007},
  year={2024},
  publisher={Wiley Online Library}
}

@article{soldozy2019biophysical,
  title={The biophysical role of hemodynamics in the pathogenesis of cerebral aneurysm formation and rupture},
  author={Soldozy, Sherif and Norat, Paul and Elsarrag, Mohamed and Chatrath, Aanan and Costello, John S. and Sokolowski, Jacob D. and Tvrdik, Petr and Kalani, M. Yashar S. and Park, Min S.},
  journal={Neurosurgical Focus},
  volume={47},
  number={1},
  pages={E11},
  year={2019},
}

@article{ngoepe2018thrombosis,
  title={Thrombosis in cerebral aneurysms and the computational modeling thereof: a review},
  author={Ngoepe, Malebogo N and Frangi, Alejandro F and Byrne, James V and Ventikos, Yiannis},
  journal={Frontiers in Physiology},
  volume={9},
  pages={306},
  year={2018},
  publisher={Frontiers Media SA}
}

@article{won2025efficacy,
  title={Efficacy and safety of a novel flow-disruptor device in a rabbit aneurysm model: a preliminary study},
  author={Won, Dong-Sung and Seong, Eunyeong and Kim, Mi Hyeon and Hwang, Sungbin and Park, Jung-Hoon and Lee, Hey-Jin and Won, Chanhee and Lee, Deok Hee},
  journal={Scientific Reports},
  volume={15},
  number={1},
  pages={31379},
  year={2025},
  publisher={Nature Publishing Group UK London}
}

@article{hecker2025artisse,
  title={The Artisse intrasaccular device: a new intrasaccular flow diverter for the treatment of cerebral aneurysms},
  author={Hecker, Constantin and Hufnagl, Clemens and Oellerer, Andreas and Griessenauer, Christoph J and Killer-Oberpfalzer, Monika},
  journal={American Journal of Neuroradiology},
  volume={46},
  number={1},
  pages={84--89},
  year={2025},
  publisher={American Journal of Neuroradiology}
}

@article{hecker2026artisse,
  title={The Artisse intrasaccular device for the treatment of cerebral aneurysms: initial experience from three Austrian neurovascular centers},
  author={Hecker, Constantin and Janu, Andrea and Pfaff, Johannes Alex Rolf and Pikija, Slaven and Sonnberger, Michael and Pangratz-Daller, Cornelia and Kral, Michael and Lunzer, Manuel and Griessenauer, Christoph J and Killer-Oberpfalzer, Monika},
  journal={Journal of NeuroInterventional Surgery},
  volume={18},
  number={1},
  pages={34--40},
  year={2026},
  publisher={British Medical Journal Publishing Group}
}

@article{zoppo2024novel,
  title={A novel intrasaccular aneurysm device with high complete occlusion rate: initial results in a rabbit model},
  author={Zoppo, Christopher T and Kolstad, Josephine W and King, Robert M and Wolfe, Thomas and Kraitem, Afif and Vardar, Zeynep and Badruddin, Aamir and Pereira, Edgard and Guerrero, Boris Pab{\'o}n and Rosqueta, Arturo S and others},
  journal={Journal of neurointerventional surgery},
  volume={16},
  number={9},
  pages={928--933},
  year={2024},
  publisher={British Medical Journal Publishing Group}
}

@misc{Voss2025,
      title={A Novel Helical Thin-Film Flow Diverter: Design, Fabrication, and Computational Assessment of Hemodynamic Performance}, 
      author={Samuel Voss and Philipp Berg and Janneck Stahl and Daniel Behme and Gabor Janiga and Rodrigo Lima de Miranda and Eckhard Quandt and Prasanth Velvaluri},
      year={2025},
      eprint={2510.05320},
      archivePrefix={arXiv},
      primaryClass={physics.med-ph},
}

@article{howard2019comprehensive,
  title={Comprehensive review of imaging of intracranial aneurysms and angiographically negative subarachnoid hemorrhage},
  author={Howard, Brian M. and Hu, Rong and Barrow, John W. and Barrow, Daniel L.},
  journal={Neurosurgical Focus},
  volume={47},
  number={1},
  pages={E20},
  year={2019},
}

@article{adams1994seeded,
  title={Seeded region growing},
  author={Adams, Rolf and Bischof, Leanne},
  journal={IEEE Transactions on Pattern Analysis and Machine Intelligence},
  volume={16},
  number={6},
  pages={641--647},
  year={1994},
}

@article{lorensen1987marching,
  title={Marching cubes: A high resolution 3D surface construction algorithm},
  author={Lorensen, William E. and Cline, Harvey E.},
  journal={ACM SIGGRAPH Computer Graphics},
  volume={21},
  number={4},
  pages={163--169},
  year={1987},
}

@article{field1988laplacian,
  title={Laplacian smoothing and Delaunay triangulations},
  author={Field, David A.},
  journal={Communications in Applied Numerical Methods},
  volume={4},
  number={6},
  pages={709--712},
  year={1988},
}

@article{ma2012computer,
  title={Computer modeling of deployment and mechanical expansion of neurovascular flow diverter in patient-specific intracranial aneurysms},
  author={Ma, Ding and Dargush, Gary F and Natarajan, Sabareesh K and Levy, Elad I and Siddiqui, Adnan H and Meng, Hui},
  journal={Journal of Biomechanics},
  volume={45},
  number={13},
  pages={2256--2263},
  year={2012},
  publisher={Elsevier}
}

@article{zhang2016towards,
  title={Towards the patient-specific design of flow diverters made from helix-like wires: an optimization study},
  author={Zhang, Mingzi and Anzai, Hitomi and Chopard, Bastien and Ohta, Makoto},
  journal={Biomedical Engineering Online},
  volume={15},
  pages={371--382},
  year={2016},
  publisher={Springer}
}

@article{reymond2025novel,
  title={A novel method for brain aneurysms computed fluid dynamics analysis after flow diverter stent implantation based on micro-computed tomography reconstruction},
  author={Reymond, Philippe and Bernava, Gianmarco and Brina, Olivier and Hofmeister, Jeremy and Rosi, Andrea and L{\"o}vblad, Karl-Olof and Machi, Paolo},
  journal={Journal of Biomechanics},
  pages={112634},
  year={2025},
  publisher={Elsevier}
}

@article{shah2021volume,
  title={Volume-based sizing of the Woven EndoBridge (WEB) device: a preliminary assessment of a novel method for device size selection},
  author={Shah, Kevin A and White, Timothy G and Teron, Ina and Link, Thomas and Dehdashti, Amir R and Katz, Jeffrey M and Woo, Henry H},
  journal={Interventional Neuroradiology},
  volume={27},
  number={4},
  pages={473--480},
  year={2021},
  publisher={SAGE Publications Sage UK: London, England}
}

@article{munoz2024modification,
  title={Modification of woven endo-bridge after intracranial aneurysm treatment: a methodology for three-dimensional analysis of shape and relative position changes},
  author={Mu{\~n}oz, Romina and Dazeo, Nicol{\'a}s and Estevez-Areco, Santiago and Janot, Kevin and Narata, Ana Paula and Rouchaud, Aymeric and Larrabide, Ignacio},
  journal={Annals of Biomedical Engineering},
  volume={52},
  number={5},
  pages={1403--1414},
  year={2024},
  publisher={Springer}
}

@article{aghli2021image,
  title={Image-based computational hemodynamic analysis of an anterior communicating aneurysm treated with the Woven EndoBridge device},
  author={Aghli, Yasaman and Dayyani, Mojtaba and Golparvar, Behzad and Baharvahdat, Humain and Blanc, Raphael and Piotin, Michel and Niazmand, Hamid},
  journal={Interdisciplinary Neurosurgery},
  volume={25},
  pages={101251},
  year={2021},
  publisher={Elsevier}
}

@article{lyu2024treatment,
  title={Treatment for middle cerebral artery bifurcation aneurysms: in silico comparison of the novel Contour device and conventional flow-diverters},
  author={Lyu, Mengzhe and Torii, Ryo and Liang, Ce and Peach, Thomas W and Bhogal, Pervinder and Makalanda, Levansri and Li, Qiaoqiao and Ventikos, Yiannis and Chen, Duanduan},
  journal={Biomechanics and Modeling in Mechanobiology},
  volume={23},
  number={4},
  pages={1149--1160},
  year={2024},
  publisher={Springer}
}

@book{antman2005nonlinear,
  author    = {Antman, Stuart S.},
  title     = {Nonlinear Problems of Elasticity},
  edition   = {2},
  series    = {Applied Mathematical Sciences},
  volume    = {107},
  publisher = {Springer},
  address   = {New York},
  year      = {2005},
}

@article{simo1986finite,
  author  = {Simo, J. C. and Vu-Quoc, L.},
  title   = {A Three-Dimensional Finite-Strain Rod Model. Part II: Computational Aspects},
  journal = {Computer Methods in Applied Mechanics and Engineering},
  volume  = {58},
  number  = {1},
  pages   = {79--116},
  year    = {1986},
}

@article{jelenic1999objectivity,
  author  = {Jeleni{\'c}, G. and Crisfield, M. A.},
  title   = {Objectivity of Strain Measures in the Geometrically Exact Three-Dimensional Beam Theory and Its Finite-Element Implementation},
  journal = {Proceedings of the Royal Society of London. Series A: Mathematical, Physical and Engineering Sciences},
  volume  = {455},
  number  = {1983},
  pages   = {1125--1147},
  year    = {1999},
}

@article{meier2015locking,
  title={A locking-free finite element formulation and reduced models for geometrically exact Kirchhoff rods},
  author={Meier, Christoph and Popp, Alexander and Wall, Wolfgang A},
  journal={Computer Methods in Applied Mechanics and Engineering},
  volume={290},
  pages={314--341},
  year={2015},
  publisher={Elsevier}
}

@article{meier2014objective,
  title={An objective 3D large deformation finite element formulation for geometrically exact curved Kirchhoff rods},
  author={Meier, Christoph and Popp, Alexander and Wall, Wolfgang A},
  journal={Computer Methods in Applied Mechanics and Engineering},
  volume={278},
  pages={445--478},
  year={2014},
  publisher={Elsevier}
}

@article{meier2019geometrically,
  title={Geometrically exact finite element formulations for slender beams: Kirchhoff--Love theory versus Simo--Reissner theory},
  author={Meier, Christoph and Popp, Alexander and Wall, Wolfgang A},
  journal={Archives of Computational Methods in Engineering},
  volume={26},
  number={1},
  pages={163--243},
  year={2019},
  publisher={Springer}
}

@article{kim2008mechanical,
  title={Mechanical modeling of self-expandable stent fabricated using braiding technology},
  author={Kim, Ju Hyun and Kang, Tae Jin and Yu, Woong-Ryeol},
  journal={Journal of Biomechanics},
  volume={41},
  number={15},
  pages={3202--3212},
  year={2008},
  publisher={Elsevier}
}

@Article{steinbrecher2020,
  author   = {Steinbrecher, Ivo and Mayr, Matthias and Grill, Maximilian J. and Kremheller, Johannes and Meier, Christoph and Popp, Alexander},
  journal  = {Computational Mechanics},
  title    = {A mortar-type finite element approach for embedding {1D} beams into {3D} solid volumes},
  year     = {2020},
  issn     = {1432-0924},
  number   = {6},
  pages    = {1377--1398},
  volume   = {66},
}

@Article{steinbrecher2022,
  author   = {Steinbrecher, Ivo and Popp, Alexander and Meier, Christoph},
  journal  = {Computational Mechanics},
  title    = {Consistent coupling of positions and rotations for embedding {1D} {C}osserat beams into {3D} solid volumes},
  year     = {2022},
  issn     = {1432-0924},
  number   = {3},
  pages    = {701--732},
  volume   = {69},
}

@article{steinbrecher2025consistent,
  title={A consistent mixed-dimensional coupling approach for 1D Cosserat beams and 2D surfaces in 3D space},
  author={Steinbrecher, Ivo and Hagmeyer, Nora and Meier, Christoph and Popp, Alexander},
  journal={Computational Mechanics},
  pages={1--28},
  year={2025},
  publisher={Springer}
}

@article{steinbrecher2026variationally,
  title={A variationally consistent beam-to-beam point coupling formulation for geometrically exact beam theories},
  author={Steinbrecher, Ivo and Hagmeyer, Nora and Meier, Christoph and Popp, Alexander},
  journal={Acta Mechanica},
  pages={1--20},
  year={2026},
  publisher={Springer}
}

@article{datz2026influence,
  title={Influence of coronary plaque morphology on local mechanical states and associated in-stent restenosis},
  author={Datz, Janina C and Steinbrecher, Ivo and Krefting, Johannes and Engel, Leif-Christopher and Popp, Alexander and Pfaller, Martin R and Schunkert, Heribert and Wall, Wolfgang A},
  journal={Journal of Biomechanical Engineering},
  volume={148},
  number={4},
  pages={041008},
  year={2026},
  publisher={American Society of Mechanical Engineers}
}

@article{datz2025patient,
  title={Patient-specific coronary angioplasty simulations—A mixed-dimensional finite element modeling approach},
  author={Datz, Janina C and Steinbrecher, Ivo and Meier, Christoph and Hagmeyer, Nora and Engel, Leif-Christopher and Popp, Alexander and Pfaller, Martin R and Schunkert, Heribert and Wall, Wolfgang A},
  journal={Computers in biology and medicine},
  volume={189},
  pages={109914},
  year={2025},
  publisher={Elsevier}
}

@Article{Khristenko2021,
  author   = {Ustim Khristenko and Stefan Schu{\ss} and Melanie Kr{\"u}ger and Felix Schmidt and Barbara Wohlmuth and Christian Hesch},
  journal  = {Computer Methods in Applied Mechanics and Engineering},
  title    = {Multidimensional coupling: A variationally consistent approach to fiber-reinforced materials},
  year     = {2021},
  issn     = {0045-7825},
  pages    = {113869},
  volume   = {382},
}

@Article{Portillo2026,
  author       = {David Portillo and Ignacio Romero},
  journal      = {Computer Methods in Applied Mechanics and Engineering},
  title        = {Embedding structures in continua: {L}inear models and finite element discretizations},
  year         = {2026},
  issn         = {0045-7825},
  pages        = {118683},
  volume       = {451},
}

@misc{4C,
  author       = {4C},
  title        = {4C: A {C}omprehensive {M}ultiphysics {S}imulation {F}ramework},
  howpublished = {\url{https://www.4c-multiphysics.org}},
  year         = {2026},
  note         = {Accessed: 2026-02-16}
}

@misc{BeamMe,
  author       = {{BeamMe Authors}},
  title        = {{B}eam{M}e -- {A} general purpose {3D} beam finite element input generator},
  howpublished = {\url{https://beamme-py.github.io/beamme}},
  year         = {2026},
  note         = {Accessed: 18 February 2026}
}

@article{bisighini2022endobeams,
  title={EndoBeams. jl: A Julia finite element package for beam-to-surface contact problems in cardiovascular mechanics},
  author={Bisighini, Beatrice and Aguirre, Miquel and Pierrat, Baptiste and Perrin, David and Avril, St{\'e}phane},
  journal={Advances in Engineering Software},
  volume={171},
  pages={103173},
  year={2022},
  publisher={Elsevier}
}

@article{kelly2019comparison,
  title={Comparison of computational modelling techniques for braided stent analysis},
  author={Kelly, Nicola and McGrath, Donnacha J and Sweeney, Caoimhe A and Kurtenbach, Kathrin and Grogan, James A and Jockenhoevel, Stefan and O'Brien, Barry J and Bruzzi, Mark and McHugh, Peter E},
  journal={Computer Methods in Biomechanics and Biomedical Engineering},
  volume={22},
  number={16},
  pages={1334--1344},
  year={2019},
  publisher={Taylor \& Francis}
}

@article{fu2017interaction,
  title={Interaction between flow diverter and parent artery of intracranial aneurysm: a computational study},
  author={Fu, Wenyu and Xia, Qixiao},
  journal={Applied Bionics and Biomechanics},
  volume={2017},
  number={1},
  pages={3751202},
  year={2017},
  publisher={Wiley Online Library}
}

@article{shanahan2017looped,
  title={Looped ends versus open ends braided stent: A comparison of the mechanical behaviour using analytical and numerical methods},
  author={Shanahan, Camelia and Tiernan, Peter and Tofail, Syed AM},
  journal={Journal of the Mechanical Behavior of Biomedical Materials},
  volume={75},
  pages={581--591},
  year={2017},
  publisher={Elsevier}
}

@article{do2024numerical,
  title={Numerical simulation of individualized flow diversion cerebral aneurysms treatment},
  author={Do, Huy Quang and Makvandi, Resam and Ding, Andreas and Juhre, Daniel},
  journal={PAMM},
  volume={24},
  number={4},
  pages={e202400209},
  year={2024},
  publisher={Wiley Online Library}
}

@article{chavalla2019simulation,
  title={Simulation of NiTi stent deployment in a realistic patient carotid artery using isogeometric analysis},
  author={Chavalla, Sharath and Hoffmann, Thomas and Juhre, Daniel},
  journal={Procedia Structural Integrity},
  volume={15},
  pages={8--15},
  year={2019},
  publisher={Elsevier}
}

@article{de2009virtual,
  title={Virtual optimization of self-expandable braided wire stents},
  author={De Beule, Matthieu and Van Cauter, Sofie and Mortier, Peter and Van Loo, Denis and Van Impe, Rudy and Verdonck, Pascal and Verhegghe, Benedict},
  journal={Medical Engineering \& Physics},
  volume={31},
  number={4},
  pages={448--453},
  year={2009},
  publisher={Elsevier}
}

@article{abdollahi2024virtual,
  title={Virtual and analytical self-expandable braided stent treatment models},
  author={Abdollahi, Reza and Shahi, Amirali and Roy, Daniel and Lessard, Simon and Mongrain, Rosaire and Soulez, Gilles},
  journal={Medical Engineering \& Physics},
  volume={126},
  pages={104145},
  year={2024},
  publisher={Elsevier}
}

@article{suzuki2017relationships,
  title={Relationships between geometrical parameters and mechanical properties for a helical braided flow diverter stent},
  author={Suzuki, Takashi and Takao, Hiroyuki and Fujimura, Soichiro and Dahmani, Chihebeddine and Ishibashi, Toshihiro and Mamori, Hiroya and Fukushima, Naoya and Murayama, Yuichi and Yamamoto, Makoto},
  journal={Technology and Health Care},
  volume={25},
  number={4},
  pages={611--623},
  year={2017},
  publisher={IOS Press}
}

@article{qiu2022influence,
  title={Influence of geometric parameters on partial compressive force and pushing performance of flow diverter},
  author={Qiu, Xiaojian and Gu, Xuelian and Liu, Chenyang and Tian, Hao and Chen, Ruina and Li, Yuanyuan},
  journal={International Journal for Numerical Methods in Biomedical Engineering},
  volume={38},
  number={2},
  pages={e3553},
  year={2022},
  publisher={Wiley Online Library}
}

@article{shang2023bending,
  title={Bending stiffness characterization of braided stent using spring-based theoretical formula},
  author={Shang, Zufeng and Ma, Jiayao},
  journal={Archive of Applied Mechanics},
  volume={93},
  number={3},
  pages={947--960},
  year={2023},
  publisher={Springer}
}

@article{zaccaria2021analytical,
  title={Analytical methods for braided stents design and comparison with FEA},
  author={Zaccaria, Alissa and Pennati, Giancarlo and Petrini, Lorenza},
  journal={Journal of the Mechanical Behavior of Biomedical Materials},
  volume={119},
  pages={104560},
  year={2021},
  publisher={Elsevier}
}

@article{alherz2016numerical,
  title={A numerical framework for the mechanical analysis of dual-layer stents in intracranial aneurysm treatment},
  author={Alherz, Ali I and Tanweer, Omar and Flamini, Vittoria},
  journal={Journal of Biomechanics},
  volume={49},
  number={12},
  pages={2420--2427},
  year={2016},
  publisher={Elsevier}
}

@article{spitz2024assessment,
  title={Assessment of intracranial aneurysm neck deformation after contour deployment},
  author={Spitz, Lena and Korte, Jana and Gaidzik, Franziska and Larsen, Naomi and Preim, Bernhard and Saalfeld, Sylvia},
  journal={International Journal of Computer Assisted Radiology and Surgery},
  volume={19},
  number={12},
  pages={2321--2327},
  year={2024},
  publisher={Springer}
}

@article{duerig1999nitinol,
  author  = {Duerig, T. and Pelton, A. and St{\"o}ckel, D.},
  title   = {An Overview of Nitinol Medical Applications},
  journal = {Materials Science and Engineering: A},
  volume  = {273--275},
  pages   = {149--160},
  year    = {1999},
}

@article{OTSUKA2005,
title = {Physical metallurgy of Ti--Ni-based shape memory alloys},
author = {Kazuhiro, Otsuka and Xiaobing, Ren},
journal = {Progress in Materials Science},
volume = {50},
number = {5},
pages = {511-678},
year = {2005},
}

@article{pelton2000optimization,
  author  = {Pelton, A. R. and DiCello, J. and Miyazaki, S.},
  title   = {Optimisation of Processing and Properties of Medical Grade Nitinol Wire},
  journal = {Minimally Invasive Therapy \& Allied Technologies},
  volume  = {9},
  number  = {1},
  pages   = {107--118},
  year    = {2000},
}

@article{duerig2000overview,
  title={An overview of superelastic stent design},
  author={Duerig, Thomas W. and Tolomeo, David E. and Wholey, Mark},
  journal={Minimally Invasive Therapy \& Allied Technologies},
  volume={9},
  number={3-4},
  pages={235--246},
  year={2000},
  publisher={Taylor \& Francis}
}

@article{pelton2022prestrain,
  title={Pre-strain and mean strain effects on the fatigue behavior of superelastic Nitinol medical devices},
  author={Pelton, Alan R. and Berg, Brian T. and Saffari, P. and Stebner, Aaron P. and Bucsek, Ashley N.},
  journal={Shape Memory and Superelasticity},
  volume={8},
  pages={64--84},
  year={2022},
  publisher={ASM International}
}

@Article{ma16196480,
TITLE = {Effect of Heat Treatment Time and Temperature on the Microstructure and Shape Memory Properties of Nitinol Wires},
AUTHOR = {Agarwal, Neha and Ryan Murphy, Josephine and Hashemi, Tina Sadat and Mossop, Theo and O'Neill, Darragh and Power, John and Shayegh, Ali and Brabazon, Dermot},
JOURNAL = {Materials},
VOLUME = {16},
YEAR = {2023},
NUMBER = {19},
publisher={MDPI}
}

@article{HWANG1983381,
title = {Compositional dependence of transformation temperatures in ternary TiNiAl and TiNiFe alloys},
author= {Hwang, Chong Moon and Wayman, Clarence Marvin},
journal = {Scripta Metallurgica},
volume = {17},
number = {3},
pages = {381-384},
year = {1983},
publisher = {Elsevier}
}

@article{Labu,
author = {Bumke, Lars and Chluba, Christoph and Ossmer, Hinnerk and Zamponi, Christiane and Kohl, Manfred and Quandt, Eckhard},
title = {Cobalt Gradient Evolution in Sputtered TiNiCuCo Films for Elastocaloric Cooling},
journal = {physica status solidi (b)},
volume = {255},
number = {2},
pages = {1700299},
year = {2018},
publisher={Wiley}
}

@article{GALL20024643,
title = {Cyclic deformation mechanisms in precipitated NiTi shape memory alloys},
journal = {Acta Materialia},
volume = {50},
number = {18},
pages = {4643-4657},
year = {2002},
author = {Gall, Ken and Maier, Hans J.},
publisher= {Elsevier}
}

@incollection{elsisy2020materials,
  title={Materials properties and manufacturing processes of nitinol endovascular devices},
  author={Elsisy, Moataz and Chun, Youngjae},
  editor={B{\'a}rtolo, Paulo Jorge and Bidanda, Bopaya},
  booktitle={Bio-materials and prototyping applications in medicine},
  pages={59--79},
  year={2020},
  publisher={Springer},
  address={Cham}
}

@article{Pelton2008,
  title   = {Fatigue and durability of Nitinol stents},
  author  = {Pelton, Alan R. and Schroeder, Veronique and Mitchell, Michael R. and Gong, Xiao-Yan and Barney, Michael and Robertson, Scott W.},
  journal = {Journal of the Mechanical Behavior of Biomedical Materials},
  volume  = {1},
  number  = {2},
  pages   = {153--164},
  year    = {2008},
}

@article{auricchio1997shape,
  title={Shape-memory alloys: macromodelling and numerical simulations of the superelastic behavior},
  author={Auricchio, Ferdinando and Taylor, Robert L and Lubliner, Jacob and others},
  journal={Computer Methods in Applied Mechanics and Engineering},
  volume={146},
  number={3-4},
  pages={281--312},
  year={1997},
  publisher={Citeseer}
}

@article{Bernini,
    author = {Bernini, Martina AND Hellmuth, Rudolf AND Dunlop, Craig AND Ronan, William AND Vaughan, Ted J.},
    journal = {PLOS ONE},
    publisher = {Public Library of Science},
    title = {Recommendations for finite element modelling of nickel-titanium stents - Verification and validation activities},
    year = {2023},
    month = {08},
    volume = {18},
    pages = {1-34},
    number = {8},

}

@article{weafer2013micro,
  title={The micro-macroscale correlation of NiTi mechanical behavior: a finite element analysis},
  author={Weafer, FM and Bruzzi, MS},
  journal={WIT Transactions on Engineering Sciences},
  volume={77},
  pages={17--30},
  year={2013},
  publisher={WIT Press}
}

@article{kutbay2023quantitative,
  title={Quantitative radial force measurements of Woven EndoBridge devices},
  author={Kutbay, Ugurhan and Algin, Oktay},
  journal={Interventional Neuroradiology},
  pages={15910199231209072},
  year={2023},
  publisher={SAGE Publications Sage UK: London, England}
}

@article{giuliodori2021numerical,
  title={Numerical modeling of bare and polymer-covered braided stents using torsional and tensile springs connectors},
  author={Giuliodori, Agustina and Hern{\'a}ndez, Joaqu{\'\i}n A and Fernandez-Sanchez, David and Galve, I{\~n}aki and Soudah, Eduardo},
  journal={Journal of Biomechanics},
  volume={123},
  pages={110459},
  year={2021},
  publisher={Elsevier}
}

@article{mckenna2021finite,
  title={A finite element investigation on design parameters of bare and polymer-covered self-expanding wire braided stents},
  author={McKenna, Ciara G and Vaughan, Ted J},
  journal={Journal of the Mechanical Behavior of Biomedical Materials},
  volume={115},
  pages={104305},
  year={2021},
  publisher={Elsevier}
}

@article{hu2024modified,
  title={Modified Theoretical Model Predicts Radial Support Capacity of Polymer Braided Stents},
  author={Hu, Xue and Liu, Qingwei and Chen, Li and Cheng, Jie and Liu, Muqing and Wu, Gensheng and Sun, Renhua and Zhao, Gutian and Yang, Juekuan and Ni, Zhonghua},
  journal={Computer Methods and Programs in Biomedicine},
  volume={246},
  pages={108063},
  year={2024},
  publisher={Elsevier}
}

@article{ubachs2023computational,
  title={Computational modeling of braided venous stents - effect of design features and device-tissue interaction on stent performance},
  author={Ubachs, Ren{\'e} and van der Sluis, Olaf and Smith, Scott and Mertens, Jake},
  journal={Journal of the Mechanical Behavior of Biomedical Materials},
  volume={142},
  pages={105857},
  year={2023},
  publisher={Elsevier}
}

@article{zhao2024finite,
  title={Finite element analysis of braided dense-mesh stents for carotid artery stenosis},
  author={Zhao, Yunchuan and Cui, Haipo},
  journal={Computer Methods in Biomechanics and Biomedical Engineering},
  volume={27},
  number={5},
  pages={609--619},
  year={2024},
  publisher={Taylor \& Francis}
}

@article{zhao2025mechanical,
  title={Mechanical characteristics of braided composite stents for carotid artery stenosis using finite element method},
  author={Zhao, Gaiping and Zhang, Zhengnan and Chen, Eryun and Ding, Fenghua and Yuan, Ruosen and Song, Chengli and Yan, Wentao and Wu, Kunneng and Wu, Jie},
  journal={Computer Methods in Biomechanics and Biomedical Engineering},
  pages={1--9},
  year={2025},
  publisher={Taylor \& Francis}
}

@article{kono2013hemodynamics,
  title={Hemodynamics of 8 different configurations of stenting for bifurcation aneurysms},
  author={Kono, K and Terada, T},
  journal={American Journal of Neuroradiology},
  volume={34},
  number={10},
  pages={1980--1986},
  year={2013},
  publisher={American Journal of Neuroradiology}
}

@article{takizawa2012comparative,
  title={A comparative study based on patient-specific fluid-structure interaction modeling of cerebral aneurysms},
  author={Takizawa, Kenji and Brummer, Tyler and Tezduyar, Tayfun E and Chen, Peng R},
  journal={Journal of Applied Mechanics},
  volume={79},
  pages={010908},
  year={2012},
  publisher={American Society of Mechanical Engineers}
}

@article{torii2009fluid,
  title={Fluid--structure interaction modeling of blood flow and cerebral aneurysm: significance of artery and aneurysm shapes},
  author={Torii, Ryo and Oshima, Marie and Kobayashi, Toshio and Takagi, Kiyoshi and Tezduyar, Tayfun E},
  journal={Computer Methods in Applied Mechanics and Engineering},
  volume={198},
  number={45-46},
  pages={3613--3621},
  year={2009},
  publisher={Elsevier}
}

@article{eken2017parallel,
  title={A parallel monolithic approach for fluid-structure interaction in a cerebral aneurysm},
  author={Eken, Ali and Sahin, Mehmet},
  journal={Computers \& Fluids},
  volume={153},
  pages={61--75},
  year={2017},
  publisher={Elsevier}
}

@article{abdehkakha2021cerebral,
  title={Cerebral aneurysm flow diverter modeled as a thin inhomogeneous porous medium in hemodynamic simulations},
  author={Abdehkakha, Armin and Hammond, Adam L and Patel, Tatsat R and Siddiqui, Adnan H and Dargush, Gary F and Meng, Hui},
  journal={Computers in Biology and Medicine},
  volume={139},
  pages={104988},
  year={2021},
  publisher={Elsevier}
}

@article{berod2022heterogeneous,
  title={A heterogeneous model of endovascular devices for the treatment of intracranial aneurysms},
  author={Berod, Alain and Chnafa, Christophe and Mendez, Simon and Nicoud, Franck},
  journal={International Journal for Numerical Methods in Biomedical Engineering},
  volume={38},
  number={2},
  pages={e3552},
  year={2022},
  publisher={Wiley Online Library}
}

@article{hagmeyer2022oneway,
  author  = {Hagmeyer, Nora and Mayr, Matthias and Steinbrecher, Ivo and Popp, Alexander},
  title   = {One-way coupled fluid--beam interaction: capturing the effect of embedded slender bodies on global fluid flow and vice versa},
  journal = {Advanced Modeling and Simulation in Engineering Sciences},
  volume  = {9},
  number  = {1},
  pages   = {9},
  year    = {2022},
}

@article{hagmeyer2024fullycoupled,
  author  = {Hagmeyer, Nora and Mayr, Matthias and Popp, Alexander},
  title   = {A fully coupled regularized mortar-type finite element approach for embedding one-dimensional fibers into three-dimensional fluid flow},
  journal = {International Journal for Numerical Methods in Engineering},
  volume  = {125},
  number  = {8},
  pages   = {e7435},
  year    = {2024},
}

@article{goetz2024anxplore,
  title={AnXplore: a comprehensive fluid-structure interaction study of 101 intracranial aneurysms},
  author={Goetz, Aur{\`e}le and Jeken-Rico, Pablo and Pelissier, Ugo and Chau, Yves and S{\'e}dat, Jacques and Hachem, Elie},
  journal={Frontiers in Bioengineering and Biotechnology},
  volume={12},
  pages={1433811},
  year={2024},
  publisher={Frontiers Media SA}
}

@article{shidhore2023comparative,
  title={Comparative Assessment of Biomechanical Parameters in Subjects With Multiple Cerebral Aneurysms Using Fluid--Structure Interaction Simulations},
  author={Shidhore, Tanmay C and Cohen-Gadol, Aaron A and Rayz, Vitaliy L and Christov, Ivan C},
  journal={Journal of Biomechanical Engineering},
  volume={145},
  number={5},
  pages={051003},
  year={2023},
  publisher={ASME},
}

@article{jeken2025investigating,
  title={Investigating Delayed Rupture of Flow Diverter-Treated Giant Aneurysm Using Simulated Fluid--Structure Interactions},
  author={Jeken-Rico, Pablo and Chau, Yves and Goetz, Aur{\`e}le and Sedat, Jacques and Hachem, Elie},
  journal={Bioengineering},
  volume={12},
  number={3},
  pages={305},
  year={2025},
  publisher={MDPI}
}

@article{campo2015review,
  title={A review of computational hemodynamics in middle cerebral aneurysms and rheological models for blood flow},
  author={Campo-Dea{\~n}o, Laura and Oliveira, M{\'o}nica SN and Pinho, Fernando T},
  journal={Applied Mechanics Reviews},
  volume={67},
  number={3},
  pages={030801},
  year={2015},
  publisher={American Society of Mechanical Engineers}
}

@article{saqr2020computational,
  title={Computational fluid dynamics simulations of cerebral aneurysm using Newtonian, power-law and quasi-mechanistic blood viscosity models},
  author={Saqr, Khalid M},
  journal={Proceedings of the Institution of Mechanical Engineers, Part H: Journal of Engineering in Medicine},
  volume={234},
  number={7},
  pages={711--719},
  year={2020},
  publisher={SAGE Publications Sage UK: London, England}
}

@article{boniforti2024endovascular,
  title={Endovascular Treatment of Intracranial Aneurysm: The Importance of the Rheological Model in Blood Flow Simulations},
  author={Boniforti, Maria Antonietta and Vittucci, Giorgia and Magini, Roberto},
  journal={Bioengineering},
  volume={11},
  number={6},
  pages={522},
  year={2024},
  publisher={MDPI}
}

@article{lampropoulos2025investigating,
  title={Investigating Hemodynamics in Intracranial Aneurysms with Irregular Morphologies: A Multiphase CFD Approach},
  author={Lampropoulos, Dimitrios S and Hadjinicolaou, Maria},
  journal={Mathematics},
  volume={13},
  number={3},
  pages={505},
  year={2025},
  publisher={MDPI}
}

@article{chowdhury2025flow,
  title={Flow Dynamics in Brain Aneurysms: A Review of Computational and Experimental Studies},
  author={Chowdhury, Prantik Roy and Lai, Victor K and Zhang, Ruihang},
  journal={Biomechanics},
  volume={5},
  number={2},
  pages={36},
  year={2025},
  publisher={MDPI}
}

@article{griessenauer2025contour,
  title={Contour Neurovascular System for endovascular embolization of cerebral aneurysms: a multicenter cohort study of 10 European neurovascular centers},
  author={Griessenauer, Christoph J and Ghozy, Sherief and Biondi, Alessandra and Hecker, Constantin and Wodarg, Fritz and Liebig, Thomas and Patankar, Tufail and Lamin, Saleh and Mart{\'\i}nez-Gald{\'a}mez, Mario and Cognard, Christophe and others},
  journal={Journal of NeuroInterventional Surgery},
  volume={17},
  number={4},
  pages={399--404},
  year={2025},
  publisher={British Medical Journal Publishing Group}
}

@article{wodarg2025embolization,
  title={Embolization of Ruptured and Unruptured Aneurysms with the Contour Neurovascular System - Summary of 106 Cases},
  author={Wodarg, Fritz and Neves, Fernando Bueno and G{\"a}rtner, Friederike and Larsen, Naomi and Peters, S{\"o}nke and Hensler, Johannes and Klintz, Tristan and Mahnke, Justus and Ahmeti, Hajrullah and Doukas, Alexander and others},
  journal={American Journal of Neuroradiology},
  volume={46},
  number={4},
  pages={698--705},
  year={2025},
  publisher={American Journal of Neuroradiology}
}

@article{bellanger2024reversed,
  title={The reversed umbrella: displacement of a Contour device into an MCA aneurysm 18 months after treatment},
  author={Bellanger, Guillaume and Darcourt, Jean and Januel, Anne-Christine and Cognard, Christophe},
  journal={Journal of NeuroInterventional Surgery},
  volume={16},
  number={2},
  pages={213--215},
  year={2024},
  publisher={British Medical Journal Publishing Group}
}

@article{velvaluri2021realistic,
  title={A realistic way to investigate the design, and mechanical properties of flow diverter stents},
  author={Velvaluri, Prasanth and Pravdivtseva, Mariya S and Hensler, Johannes and Wodarg, Fritz and Jansen, Olav and Quandt, Eckhard and H{\"o}vener, Jan-Bernd},
  journal={Expert Review of Medical Devices},
  volume={18},
  number={6},
  pages={569--579},
  year={2021},
  publisher={Taylor \& Francis}
}

@article{sindeev2018phasecontrast,
  title={Phase-contrast MRI versus numerical simulation to quantify hemodynamical changes in cerebral aneurysms after flow diverter treatment},
  author={Sindeev, S. and Arnold, P. G. and Frolov, S. and Prothmann, S. and Liepsch, D. and Balasso, A. and Berg, Philipp and Kaczmarz, S. and Kirschke, J. S.},
  journal={PLOS ONE},
  volume={13},
  number={1},
  pages={e0190696},
  year={2018},
}

@article{Hadad2023,
author = {Hadad, Sara and Karnam, Yogesh and Mut, Fernando and Lohner, Rainald and Robertson, Anne M and Kaneko, Naoki and Cebral, Juan R},
title = {Computational fluid dynamics-based virtual angiograms for the detection of flow stagnation in intracranial aneurysms.},
journal = {International Journal for Numerical Methods in Biomedical Engineering},
volume = {39},
number = {8},
pages = {e3740},
year = {2023},
publisher={Wiley}
}

@article{korte2024vitro,
  title={In vitro and in silico assessment of flow modulation after deploying the Contour Neurovascular System in intracranial aneurysm models},
  author={Korte, Jana and Gaidzik, Franziska and Larsen, Naomi and Sch{\"u}tz, Erik and Damm, Timo and Wodarg, Fritz and H{\"o}vener, Jan-Bernd and Jansen, Olav and Janiga, G{\'a}bor and Berg, Philipp and others},
  journal={Journal of NeuroInterventional Surgery},
  volume={16},
  number={8},
  pages={815--823},
  year={2024},
  publisher={BMJ Publishing Group Ltd. BMA House, Tavistock Square, London, WC1H 9JR}
}

@article{pravdivtseva20213d,
  title={3D-printed, patient-specific intracranial aneurysm models: From clinical data to flow experiments with endovascular devices},
  author={Pravdivtseva, Mariya S and Peschke, Eva and Lindner, Thomas and Wodarg, Fritz and Hensler, Johannes and Gabbert, Dominik and Voges, Inga and Berg, Philipp and Barker, Alex J and Jansen, Olav and others},
  journal={Medical Physics},
  volume={48},
  number={4},
  pages={1469--1484},
  year={2021},
  publisher={Wiley Online Library}
}

@article{bisighini2022fabrication,
  title={Fabrication of compliant and transparent hollow cerebral vascular phantoms for in vitro studies using 3D printing and spin--dip coating},
  author={Bisighini, Beatrice and Di Giovanni, Pierluigi and Scerrati, Alba and Trovalusci, Federica and Vesco, Silvia},
  journal={Materials},
  volume={16},
  number={1},
  pages={166},
  year={2022},
  publisher={MDPI}
}

@article{berg2019review,
  title={A review on the reliability of hemodynamic modeling in intracranial aneurysms: why computational fluid dynamics alone cannot solve the equation},
  author={Berg, Philipp and Saalfeld, Sylvia and Vo{\ss}, Samuel and Beuing, Oliver and Janiga, G{\'a}bor},
  journal={Neurosurgical Focus},
  volume={47},
  number={1},
  pages={E15},
  year={2019},
  publisher={American Association of Neurological Surgeons}
}

@article{Voss2019,
  title   = {Multiple Aneurysms AnaTomy CHallenge 2018 (MATCH) - Phase Ib: Effect of morphology on hemodynamics},
  author  = {Vo{\ss}, Samuel and Beuing, Oliver and Janiga, G{\'a}bor and Berg, Philipp},
  journal = {PLOS ONE},
  volume  = {14},
  number  = {5},
  pages   = {e0216813},
  year    = {2019},
   publisher = {Public Library of Science (PLOS)},
}

@article{Paritala2023,
  title   = {Reproducibility of the computational fluid dynamic analysis of a cerebral aneurysm monitored over a decade},
  author  = {Paritala, Phani Kumari and Anbananthan, Haveena and Hautaniemi, Jacob and Smith, Macauley and George, Antony and Allenby, Mark and Benitez Mendieta, Jessica and Wang, Jiaqiu and Maclachlan, Liam and Liang, Ee Shern and Prior, Marita and Yarlagadda, Prasad K. D. V. and Winter, Craig and Li, Zhiyong},
  journal = {Scientific Reports},
  volume  = {13},
  pages   = {219},
  year    = {2023},
   publisher = {Nature},
}

@article{Yalman2025,
author = {Yalman, Alain and Jafari, Arman and L{\'e}ger, {\'E}tienne and Mastroianni, Michael-Anthony and Teimouri, Kowsar and Savoji, Houman and Collins, D. Louis and Kadem, Lyes and Xiao, Yiming},
title = {Design, manufacturing, and multi-modal imaging of stereolithography 3D printed flexible intracranial aneurysm phantoms},
journal = {Medical Physics},
volume = {52},
number = {2},
pages = {742-749},
year = {2025},
publisher={Wiley}
}

@article{karam2023,
  title={Additive manufacturing of patient-specific high-fidelity and thickness-controlled cerebral aneurysm geometries},
  author={Karam, Sandy and Shirdade, Nikhil and Madden, Benjamin and Rheinstadter, Justin and Church, Ephraim W and Brindise, Melissa C and Manogharan, Guha},
  journal={Manufacturing Letters},
  volume={35},
  pages={770--777},
  year={2023},
  publisher={Elsevier}
}

@InProceedings{Zimmermann2021,
author="Zimmermann, Judith and Baumler, Kathrin and Loecher, Michael and Cork, Tyler E. and Kolawole, Fikunwa O. and Gifford, Kyle and Marsden, Alison L. and Fleischmann, Dominik and Ennis, Daniel B.",
title="Quantitative Hemodynamics in Aortic Dissection: Comparing in Vitro MRI with FSI Simulation in a Compliant Model",
editor="Ennis, Daniel B. and Perotti, Leighton E. and Wang, Vicky Y.",
booktitle="Functional Imaging and Modeling of the Heart",
year="2021",
pages="575--586",
publisher="Springer International Publishing",
address="Cham",
}

@article{bisighini2023machine,
  title={Machine learning and reduced order modelling for the simulation of braided stent deployment},
  author={Bisighini, Beatrice and Aguirre, Miquel and Biancolini, Marco Evangelos and Trovalusci, Federica and Perrin, David and Avril, St{\'e}phane and Pierrat, Baptiste},
  journal={Frontiers in Physiology},
  volume={14},
  pages={1148540},
  year={2023},
  publisher={Frontiers Media SA}
}

@article{bisighini2023patient,
  title={Patient-specific computational modelling of endovascular treatment for intracranial aneurysms},
  author={Bisighini, Beatrice and Aguirre, Miquel and Pierrat, Baptiste and Avril, St{\'e}phane},
  journal={Brain Multiphysics},
  volume={5},
  pages={100079},
  year={2023},
  publisher={Elsevier}
}

@article{pramanik2026contact,
  title={Contact-resolved deployment of the Contour Neurovascular System in patient-specific intracranial aneurysms},
  author={Pramanik, Ratnadeep and Gie{\ss}ler, Fina and Frank, Martin and Steinbrecher, Ivo and Mayr, Matthias and Saalfeld, Sylvia and Popp, Alexander},
  journal={arXiv preprint arXiv:2607.13972},
  year={2026}
}

@article{dengiz2025thin,
  title={Thin-film NiTi intrasaccular implant with flaps for aneurysm treatments},
  author={Dengiz, D and Velvaluri, P and Grotemeyer, P and Pravdivtseva, MS and Wodarg, F and Watkinson, J and Mackensen, E and Jansen, O and Quandt, E},
  journal={Biomaterials Advances},
  volume={174},
  pages={214311},
  year={2025},
  publisher={Elsevier}
}

@article{dengiz2026mechanical,
  title={Mechanical study of NiTi intrasaccular aneurysm implants with flaps},
  author={Dengiz, Duygu and Velvaluri, Prasanth and Grotemeyer, Patrick and Jansen, Olav and Quandt, Eckhard},
  journal={Engineering Research Express},
  volume={8},
  number={8},
  pages={085507},
  year={2026},
  publisher={IOP Publishing}
}

@article{gaub2024flow,
  title={Flow diversion for endovascular treatment of intracranial aneurysms: past, present, and future directions},
  author={Gaub, Michael and Murtha, Greg and Lafuente, Molly and Webb, Matthew and Luo, Anqi and Birnbaum, Lee A and Mascitelli, Justin R and Al Saiegh, Fadi},
  journal={Journal of Clinical Medicine},
  volume={13},
  number={14},
  pages={4167},
  year={2024},
  publisher={MDPI}
}

@article{bhogal2019endosaccular,
  title={Endosaccular flow disruption: where are we now?},
  author={Bhogal, Pervinder and Udani, Sundip and Cognard, Christophe and Piotin, Michel and Brouwer, Patrick and Sourour, Nader-Antoine and Andersson, Tommy and Makalanda, Levansri and Wong, Ken and Fiorella, David and others},
  journal={Journal of NeuroInterventional Surgery},
  volume={11},
  number={10},
  pages={1024--1025},
  year={2019},
  publisher={BMJ Publishing Group Ltd. BMA House, Tavistock Square, London, WC1H 9JR}
}

@article{frank2026mechanical,
  title={Mechanical Modeling of Braided Neurovascular Flow Diverters using a Beam-to-Beam and Beam-to-Surface Contact Formulation},
  author={Frank, Martin and Steinbrecher, Ivo and Mayr, Matthias and Popp, Alexander},
  journal={arXiv preprint arXiv:2607.29446},
  year={2026}
}

@article{lespagnol2024mixed,
  title={A mixed-dimensional formulation for the simulation of slender structures immersed in an incompressible flow},
  author={Lespagnol, Fabien and Grandmont, C{\'e}line and Zunino, Paolo and Fern{\'a}ndez, Miguel A},
  journal={Computer Methods in Applied Mechanics and Engineering},
  volume={432},
  pages={117316},
  year={2024},
  publisher={Elsevier}
}

@article{konyukhov2018consistent,
  title={Consistent development of a beam-to-beam contact algorithm via the curve-to-solid beam contact: analysis for the nonfrictional case},
  author={Konyukhov, Alexander and Mrenes, Oana and Schweizerhof, Karl},
  journal={International Journal for Numerical Methods in Engineering},
  volume={113},
  number={7},
  pages={1108--1144},
  year={2018},
  publisher={Wiley Online Library}
}

@article{lalonde2017modeling,
  title={Modeling multilayered wire strands, a strategy based on 3D finite element beam-to-beam contacts-Part I: Model formulation and validation},
  author={Lalonde, S{\'e}bastien and Guilbault, Raynald and L{\'e}geron, Fr{\'e}d{\'e}ric},
  journal={International Journal of Mechanical Sciences},
  volume={126},
  pages={281--296},
  year={2017},
  publisher={Elsevier}
}

@article{bali2025finite,
  title={A finite volume adaptation of beam-to-beam contact interactions implemented for geometrically exact Simo--Reissner beams},
  author={Bali, Seevani and Tukovi{\'c}, {\v{Z}}eljko and Cardiff, Philip and Ivankovi{\'c}, Alojz and Pakrashi, Vikram},
  journal={Computational mechanics},
  volume={75},
  number={1},
  pages={237--263},
  year={2025},
  publisher={Springer}
}

@article{ait2025three,
  title={Three-dimensional beam-to-beam frictional contact with small-sliding and finite rotations},
  author={A{\"\i}t Ammar, Karim and Guidault, Pierre-Alain and Boucard, Pierre-Alain and Said, Julien and Hafid, Fikri},
  journal={Computational Mechanics},
  volume={76},
  number={4},
  pages={1135--1159},
  year={2025},
  publisher={Springer}
}

\section*{Acknowledgements}
\noindent This research is supported by the IDIR (Institute for Digital Implant Research) Project, a cooperation financed by Kiel University, University Hospital Schleswig-Holstein, and Helmholtz Zentrum Hereon. R.P. acknowledges funding for open-access publication of this manuscript by the University of the Bundeswehr Munich. M.F. and A.P. acknowledge funding by the Deutsche Forschungsgemeinschaft (SPP2311, project number: 465242983). P.V. acknowledges funding from the DFG through project number 527449568. P.B. acknowledges funding from the Federal Ministry of Research, Technology, and Space within the Forschungscampus STIMULATE (grant no. 13GW0835A) and the Deutsche Forschungsgemeinschaft (SPP2311, project number: 548907942). M.S.P. acknowledges funding from the German Federal Ministry of Research, Technology and Space (BMFTR) through the STRIVE project (project number: 01ZU2510).

\section*{Author contributions} \textbf{R.P.}: Conceptualization, data analysis, project administration, figure visualization, writing -- original draft, writing -- review and editing. \textbf{M.F.}: Figure visualization, writing -- review and editing. \textbf{D.D., M.S.P., P.V., I.S., M.M., P.B., S.S., N.L., O.J.}: Writing -- review and editing. \textbf{A.P.}: Conceptualization, funding acquisition, writing -- review and editing. All authors read and approved the final manuscript.

\section*{Data Availability}
No new data were created or analyzed in this study. Data sharing is not applicable.

\section*{Competing interests}
\noindent The authors declare no competing interests.

\end{document}